\documentclass[journal,10pt]{IEEEtran}
\IEEEoverridecommandlockouts
\usepackage{pifont}
\usepackage[top=0.7in, bottom=1in, left=0.625in, right=0.625in]{geometry}
\usepackage{textcomp}

\usepackage[T1]{fontenc}
\usepackage{graphicx}
\usepackage{subcaption}
\usepackage{colortbl}
\usepackage{xcolor}

\definecolor{rowgray}{gray}{0.95}
\definecolor{rowwhite}{gray}{1.0}
\usepackage{float}

\usepackage{cite}
\usepackage{amsmath,amssymb,amsfonts}
\usepackage{algorithmic}
\usepackage{textcomp}
\usepackage[table]{xcolor}
\usepackage{breqn}
\usepackage{algorithm,algorithmic}
\usepackage{algorithm,algorithmic,amsbsy,amsm     
 ath,amssymb,epsfig,bbm,mathrsfs,fancyhdr,fancyvrb,url}
\usepackage{amssymb}
 
\usepackage{amsthm}
\usepackage{adjustbox,lipsum}
\usepackage{amsfonts}
\usepackage{epsfig}
\usepackage{amssymb}
\usepackage{amsmath}
\usepackage{cite}

\usepackage{multirow}
\usepackage{rotating}
\usepackage{tabularx}
\usepackage{array}
\usepackage{booktabs}
\usepackage{anyfontsize}
\usepackage{color,soul}
\usepackage{blindtext}
\usepackage{authblk}

\usepackage{amsmath}
\usepackage{mathrsfs}

\usepackage{stfloats}
\usepackage{verbatim}
\usepackage{balance}
    
\def\BibTeX{{\rm B\kern-.05em{\sc i\kern-.025em b}\kern-.08em
    T\kern-.1667em\lower.7ex\hbox{E}\kern-.125emX}}

\title{HAPS-RIS and UAV Integrated Networks: \\
A Unified Joint Multi-objective Framework}
\author{Arman Azizi\thanks{
Arman Azizi, the corresponding author (e-mail: azizia@tcd.ie), and Arman Farhang are with the Department of Electronic and Electrical Engineering, Trinity College Dublin, Dublin, D02 PN40 Ireland. Mostafa Rahmani Ghourtani and Hamed Ahmadi are with the School of Physics, Engineering and Technology, University of York, YO10 5DD York, U.K. Mustafa A. Kishk is with the Department of Electronic Engineering, Maynooth University, Maynooth, W23 F2H6 Ireland. This manuscript has emanated from research conducted with the financial support of Taighde Éireann – Research
Ireland under Grant numbers 18/CRT/6222 and 13/RC/2077\textunderscore P2. For the purpose of Open Access, the author has applied a CC BY public copyright licence to any Author Accepted Manuscript version arising from this
submission.}, Mostafa Rahmani Ghourtani, Mustafa A. Kishk, Hamed Ahmadi, Arman Farhang}

\begin{document}
 \maketitle

 \begin{abstract} 
Future 6G non‑terrestrial networks aim to deliver ubiquitous connectivity to remote and undeserved regions, but unmanned aerial vehicle (UAV) base stations face fundamental challenges such as limited numbers and power budgets. To overcome these obstacles,
high‑altitude platform station (HAPS) equipped with a reconfigurable intelligent surface (RIS), so-called HAPS-RIS, is a promising candidate. We propose a novel unified joint multi-objective framework where UAVs and HAPS-RIS are fully integrated to extend coverage and enhance network performance. This joint multi-objective design maximizes the number of users served by the HAPS‑RIS, minimizes the number of UAVs deployed and minimizes the total average UAV path loss subject to quality‑of‑service (QoS) and resource constraints. We propose a novel low-complexity solution strategy by proving the equivalence between minimizing the total average UAV path loss upper bound and k-means clustering, deriving a practical closed-form RIS phase-shift design, and introducing a mapping technique that collapses the combinatorial assignments into a zone radius and a bandwidth-portioning factor. Then, we propose a dynamic Pareto optimization technique to solve the transformed optimization problem. Extensive simulation results demonstrate that the proposed framework adapts seamlessly across operating regimes. A HAPS‑RIS-only setup achieves full coverage at low data rates, but UAV assistance becomes indispensable as rate demands increase. By tuning a single bandwidth portioning factor, the model recovers UAV‑only, HAPS‑RIS‑only and equal bandwidth portioning baselines within one formulation and consistently surpasses them across diverse rate requirements. The simulations also quantify a tangible trade‑off between RIS scale and UAV deployment, enabling designers to trade increased RIS elements for fewer UAVs as service demands evolve.

\end{abstract}

\vspace{-5mm}
\section{Introduction}
\IEEEPARstart{A} central objective of sixth generation
wireless networks (6G) is to provide ubiquitous and reliable connectivity, enabling seamless connectivity for users and devices anytime and anywhere, at sustainable deployment costs \cite{3gpp2020study}. This objective explicitly encompasses extending coverage to remote regions where deploying terrestrial infrastructure is either technically infeasible or economically prohibitive \cite{3gpp2020study, yaacoub2020key}. To this end, unmanned aerial vehicles (UAVs) can serve as flexible aerial base stations, enabling rapid coverage extension in regions lacking cellular infrastructure or where terrestrial deployment is infeasible. Such scenarios include remote regions and disaster-affected areas where terrestrial communication infrastructure may be unavailable or damaged, as well as environments where the high cost of terrestrial infrastructure renders conventional network deployment impractical \cite{mozaffari2019tutorial, azizi2019profit, azizi2019joint}. However, the performance of UAV-assisted networks is fundamentally bounded by constraints on the number of available UAVs and their finite power budgets \cite{mozaffari2019tutorial, sabzehali2022optimizing}. As a consequence of these challenges, achieving full coverage becomes infeasible, leaving a portion of users without service.

\begin{figure*}[h]
\centering
\includegraphics[scale=0.44]{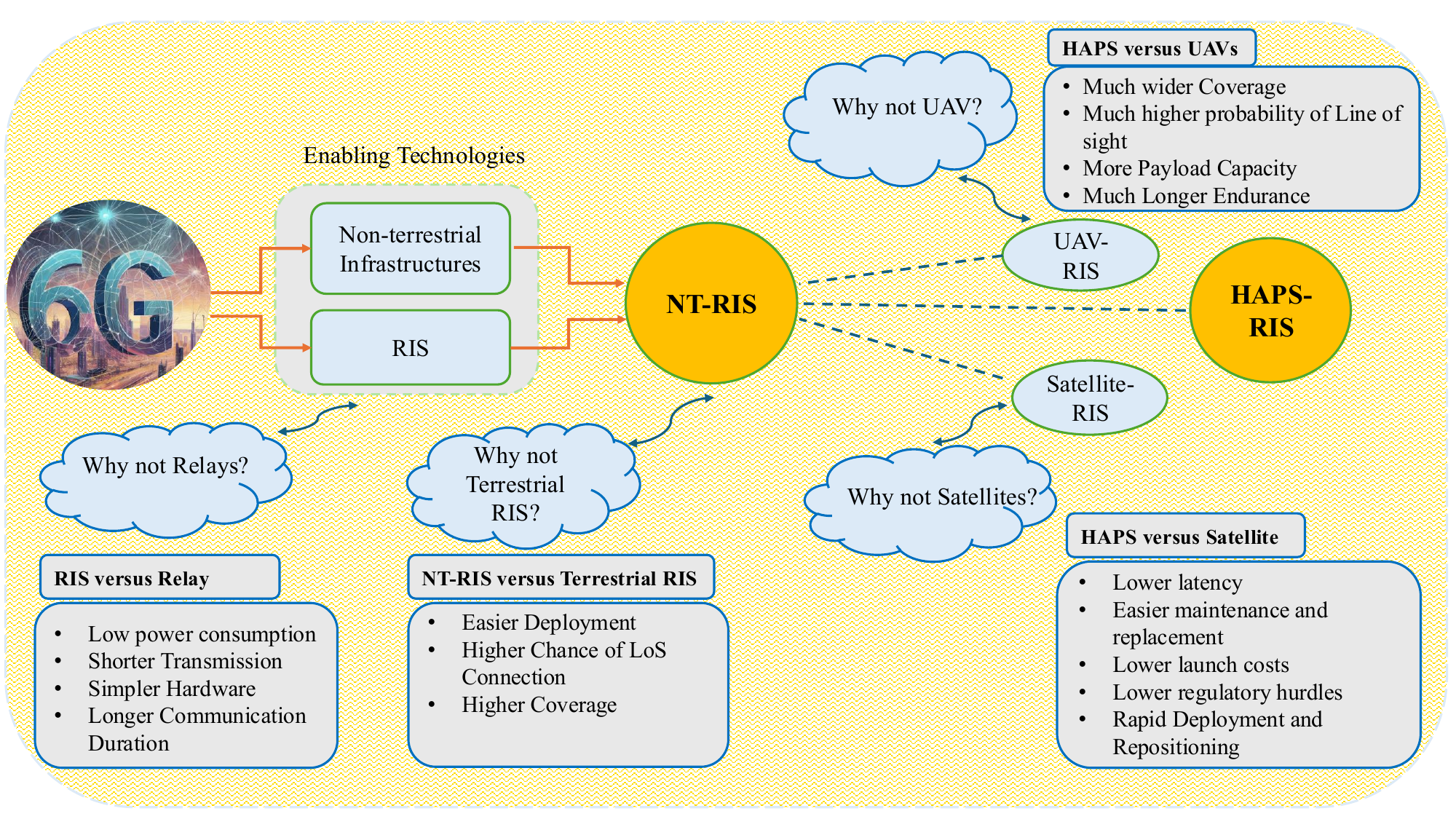}
    \caption{The motivation roadmap for choosing HAPS-RIS over alternative cases.}
    \label{key}
\end{figure*}

One approach to address these challenges is to leverage alternative non-terrestrial infrastructures as transmitters or relays.
However, this approach is neither cost-efficient nor low-complexity, as it necessitates the deployment of multiple active antennas on the non-terrestrial platform \cite{azari2022evolution, ye2022nonterrestrial}. Another alternative is to exploit existing terrestrial infrastructure, even when located at considerable distances under severe blockage conditions, to support the UAV-based network. This can be achieved by coupling them with a non-terrestrial platform equipped with a reconfigurable intelligent surface (RIS).
This architecture is referred to as non-terrestrial RIS (NT-RIS), which functions as a smart reflective layer acting as an intelligent intermediary for signal reflection \cite{ye2022nonterrestrial}. Leveraging NT-RIS offers lower cost and reduced complexity compared to the aforementioned approaches that rely on other types of non-terrestrial infrastructures \cite{ye2022nonterrestrial}. In recent years, a considerable body of research has investigated the benefits of incorporating NT-RIS into wireless networks, see \cite{ye2022nonterrestrial, ramezani2022toward, alfattani2021aerial, alfattani2021link} and references therein.
High-altitude platform station (HAPS)-RIS emerges as a particularly promising realization of NT-RIS, offering distinct advantages over alternative platforms such as satellite-RIS and UAV-RIS \cite{kurt2021vision, alfattani2021aerial, alfattani2021link}.  Fig.~\ref{key} illustrates the rationale and decision roadmap that motivate the selection of HAPS-RIS as the most suitable candidate among alternative deployment options.

Exploiting HAPS--RIS with a control station (CS), can effectively address the fundamental challenges of UAV-assisted networks, namely the limitation on the number of deployable UAVs and their limited power budgets. However, such integration inevitably introduces new resource-related challenges, among which the most critical is the limitation on the available bandwidth. As HAPS--RIS and UAVs jointly participate in serving users, efficient coordination and bandwidth coupling between the two subsystems become essential. Therefore, to effectively address both the inherent UAV limitations and the resource constraints arising from their integration, the deployment of HAPS--RIS must be considered within a unified design framework. Such a framework should jointly account for the intrinsic limitations of UAV-based access and the complementary capabilities offered by HAPS-RIS, and capture them as coupled design metrics. \textcolor{black}{Accordingly, the main research question that arises is ``\textit{How can HAPS-RIS and UAV infrastructures be jointly designed within a unified system-level multi-objective framework?}"} 

\vspace{-2mm}
\subsection{Related Work}
Motivated by the research question outlined in the previous subsection, we review the related literature by first examining HAPS--RIS–assisted networks without UAV integration, and then discussing studies that incorporate both UAVs and HAPS--RIS in non-terrestrial communication systems. Table~\ref{Comparison_Table} summarizes the taxonomy of the related literature. Early studies such as \cite{alfattani2021link} examine optimal HAPS--RIS placement by maximizing the received signal power under physical link-budget constraints. This line of research is extended in \cite{alfattani2022beyond,alfattani2023resource}, where joint optimization problems are formulated to improve throughput, fairness, and resource efficiency through power control, user association, and RIS phase-shift design. Additional works address complementary aspects, including learning-based channel estimation \cite{tekbiyik2021channel}, closed-form RIS phase-shift solutions for cascaded channel gain maximization and delay-Doppler spread mitigation \cite{azizi2023ris}, outage probability analysis under practical impairments \cite{benaya2024outage,benaya2025outage}, sum-rate and energy-efficiency trade-offs in active RIS deployments \cite{karaman2025trade}, and performance characterization under misalignment and positioning uncertainties \cite{deka2024performance}. Collectively, these studies demonstrate the potential of HAPS--RIS systems, yet they are confined to standalone HAPS--RIS architectures without UAV-assisted access.

A second category of works introduces UAVs into HAPS--RIS--enabled systems; however, in most of these studies, UAVs are treated as \emph{aerial users} rather than as part of the communication infrastructure. For example, secure communication in HAPS--RIS--assisted networks is investigated in \cite{yuan2022secure} and \cite{memarian2025enhancing}. While the former relies solely on RIS phase-shift optimization, the latter achieves secrecy rate maximization via joint BS beamforming and RIS phase-shift design.
Similarly, \cite{le2024harvested} employs a HAPS--RIS to enhance the energy harvesting performance of a single UAV, modeling the UAV as an energy-harvesting terminal and analyzing the average harvested energy under line-of-sight (LoS) and vibration constraints. From a system-level perspective, these UAV-as-user models exhibit limited structural differences from HAPS--RIS--only networks, as the UAVs do not participate in access provisioning.
The integration of UAVs as active \emph{network infrastructure} in HAPS--RIS--assisted systems remains largely unexplored. To the best of our knowledge, the only existing study in this direction is our preliminary work \cite{azizi2025exploring}, where a HAPS--RIS--assisted multi-UAV architecture is introduced and UAVs operate as aerial access nodes. While it establishes the feasibility of UAVs as infrastructure, it addresses only one challenge, namely the limitation on the number of UAVs. Other critical challenges such as UAV power-budget limitation and bandwidth limitation are left open. Moreover, \cite{azizi2025exploring} relies on a disjoint and hierarchical design, where the HAPS--RIS and UAV systems operate with dedicated bandwidth resources and are optimized sequentially, leaving joint bandwidth utilization and system-level coupling unaddressed.

\begin{table*}[htbp]
\centering
\caption{Comparison of HAPS-RIS enabled works.}
\label{Comparison_Table}
\renewcommand{\arraystretch}{1.2}
{\footnotesize
\begin{tabular}{|m{0.5cm}|m{1.7cm}|m{2.2cm}|m{3.6cm}|m{3.9cm}|m{3cm}|}
\hline
\rowcolor{gray!15}
\textbf{Ref} &
\textbf{UAV Integration Level} &
\textbf{Optimization Type} &
\textbf{Objective Function} &
\textbf{Optimization Variable} &
\textbf{Constraint} \\
\hline

\rowcolor{rowgray}
\cite{alfattani2021link} 
& ~~~~~~\ding{55} 
& Analytical 
& Max. Received signal power 
& HAPS-RIS location 
& Physical link-budget \\
\hline

\rowcolor{rowwhite}
\cite{alfattani2022beyond} 
& ~~~~~~\ding{55} 
& Joint 
& Max. Throughput, fairness, and Min. RIS-unit; separately.
& CS power, RIS parameters
& QoS, CS power, RIS-size, and phase  \\
\hline

\rowcolor{rowgray}
\cite{alfattani2023resource} 
& ~~~~~~\ding{55} 
& Joint
& Max. Resource efficiency  
& User association, CS power, and RIS phase shifts 
& QoS, CS power, HAPS size, phase shifts\\
\hline

\rowcolor{rowwhite}
\cite{tekbiyik2021channel} 
& ~~~~~~\ding{55} 
& Learning-based (Single Obj.) 
& Min. Channel estimation \newline error
& Learning model parameters
&  Pilot overhead; hardware impairments \\
\hline

\rowcolor{rowgray}
\cite{azizi2023ris} 
&  ~~~~~~\ding{55} 
&  Multi-objective (closed-form)
& Max. Channel gain; Min. Delay/Doppler spread 
&RIS phase shifts 
& Causality of phase shifts \\
\hline

\rowcolor{rowwhite}
\cite{benaya2024outage} 
& ~~~~~~\ding{55} 
& Analytical 
& Min. Outage probability 
& RIS phase shifts, system \newline parameters
& outage threshold, single-antenna nodes \\
\hline

\rowcolor{rowgray}
\cite{benaya2025outage} 
& ~~~~~~\ding{55} 
& Analytical 
& Min. Outage probability 
& RIS phase shifts, system parameters
& THz impairments, \newline outage threshold, single-antenna nodes \\
\hline

\rowcolor{rowwhite}
\cite{karaman2025trade} 
& ~~~~~~\ding{55} 
& Joint, single-objective
& Max. Sum-rate, energy-efficiency; separately.
& user transmit power, RIS parameters
& Minimum QoS per user; total CS power; HAPS size; \\
\hline

\rowcolor{rowwhite}
\cite{deka2024performance} 
& ~~~~~~\ding{55} 
& Analytical
& Outage probability and BER characterization
& RIS phase shifts; system parameters
& HAPS orientation, \newline position deviation \\
\hline

\rowcolor{rowwhite}
\cite{yuan2022secure} 
& As a user
& Single-objective 
& Max. Secrecy rate 
& RIS phase shifts
& Finite phase resolution \\
\hline

\rowcolor{rowgray}
\cite{memarian2025enhancing} 
& As a user
& Joint Single objective
& Max. Instantaneous secrecy rate 
& RIS phase shifts, beamforming vector
& BS transmit beamforming vector, phase constraints \\
\hline

\rowcolor{rowgray}
\cite{le2024harvested} 
&  As a user
& Analytical
& Evaluate average harvested energy at UAV 
& RIS size,  power parameters
& RIS vibration, \newline LoS requirement\\
\hline

\rowcolor{rowgray}
\cite{azizi2025exploring} 
& Multiple UAVs As Infrastrucrure
& Hierarchical 
& Max. HAPS-RIS users; Min. number of UAVs 
& UAV number; \newline 
zone association 
& Rate requirements \\
\hline

\rowcolor{blue!10}
\textbf{Ours} 
& Multiple UAVs as Infra.
& Joint, Multi-objective 
& Max. HAPS-RIS users; Min UAVs; Min. average UAV path loss 
& UAV location \& number; RIS phase shifts; subcarrier allocation; user association; RIS element assignment
& Rate requirements; power budgets; exclusive association; OFDMA constraints \\
\hline

\end{tabular}
}
\end{table*}
\subsection{Our Contributions}
While there has been considerable progress in exploiting HAPS-RIS to improve wireless networks,  critical gaps remain in applying this technology to UAV-assisted networks. This paper addresses the aforementioned gaps in the literature, and hence,   main contributions of this work are highlighted as follows.

\subsubsection{A Comprehensive Unified System-level Framework}
 We introduce a novel unified system-level framework for integrating HAPS-RIS and UAV integrated network by designing a novel multi-objective joint optimization framework that tightly couples HAPS-RIS and UAV-based networks in non-terrestrial networks. To the best of our knowledge, this architecture constitutes the most comprehensive system-level optimization framework for HAPS-RIS and UAV integrated networks, as benchmarked in Table~I.

\subsubsection{A Low-Complexity Solution Strategy}
We propose a novel solution strategy to reduce the original high-complexity optimization problem to a low-complexity framework through mathematical derivations provided in \emph{Theorem 1}, \emph{Lemma 1}, and \emph{Proposition 1}. Then, we solve the relaxed optimization problem with the proposed dynamic Pareto optimization approach.
\begin{itemize}
    \item  We obtain a theoretical link between classical optimization and unsupervised learning, as proven in \emph{Theorem 1}. We show that minimizing the upper bound of the total average UAV path loss is mathematically equivalent to performing $k$-means clustering as an unsupervised learning. This result allows the UAV deployment and user association problem to be solved through a clustering operation rather than optimization, effectively transforming the original three-objective formulation into a $k$-means–driven two-objective problem. 
    \item We introduce a practical RIS phase-shift design, \textit{in Lemma 1}, that leverages the geometric properties of the HAPS-RIS scenario to obtain a simple closed-form expression for the RIS phase shifts. This eliminates the need to optimize a massive number of RIS phase shifts and makes their design straightforwardly implementable.
    \item We propose a mapping technique, presented in \emph{Proposition 1}, which transforms the high-dimensional combinatorial association problem into a low-dimensional formulation using zone-based association and bandwidth portioning variables.
    \item To solve the transformed optimization problem, we propose a network-aware dynamic Pareto optimization framework. The priority structure is not arbitrarily imposed but emerges naturally from the physical and operational characteristics of the network. In particular, prioritizing HAPS–RIS coverage is shown to be essential for fully exploiting its capability before relying on UAV deployment.
    \item We obtain the computational complexity and show that the proposed framework scales polynomially with the number of users, UAVs, subcarriers, and RIS elements, confirming its scalability for large-scale HAPS-RIS deployments.

\end{itemize}

\subsubsection{System-Level Insights and Design Guidelines} Through extensive simulations, we provide rate regime design guidelines, benchmarking, and trade-offs for the proposed framework.
\begin{itemize}

\item We reveal a regime-dependent behavior in which HAPS-RIS suffices at low data rates, while higher rate regime necessitate UAV assistance.

\item We show that all conventional baseline architectures emerge as special cases of the proposed framework through a novel design parameter termed the bandwidth portioning factor, which unifies HAPS-RIS and UAV-based deployments within a single formulation. We benchmark the proposed framework against baselines across a wide range of rate requirements, consistently demonstrating its superior performance.

\item We reveal a quantifiable trade-off between UAV deployment and RIS size, where increasing RIS elements can effectively reduce the need for UAVs across different rate regimes.

\end{itemize}



   

\section{The Proposed Network Architecture}  \label{section 2}

\begin{figure}[t]
\includegraphics[scale=0.27]{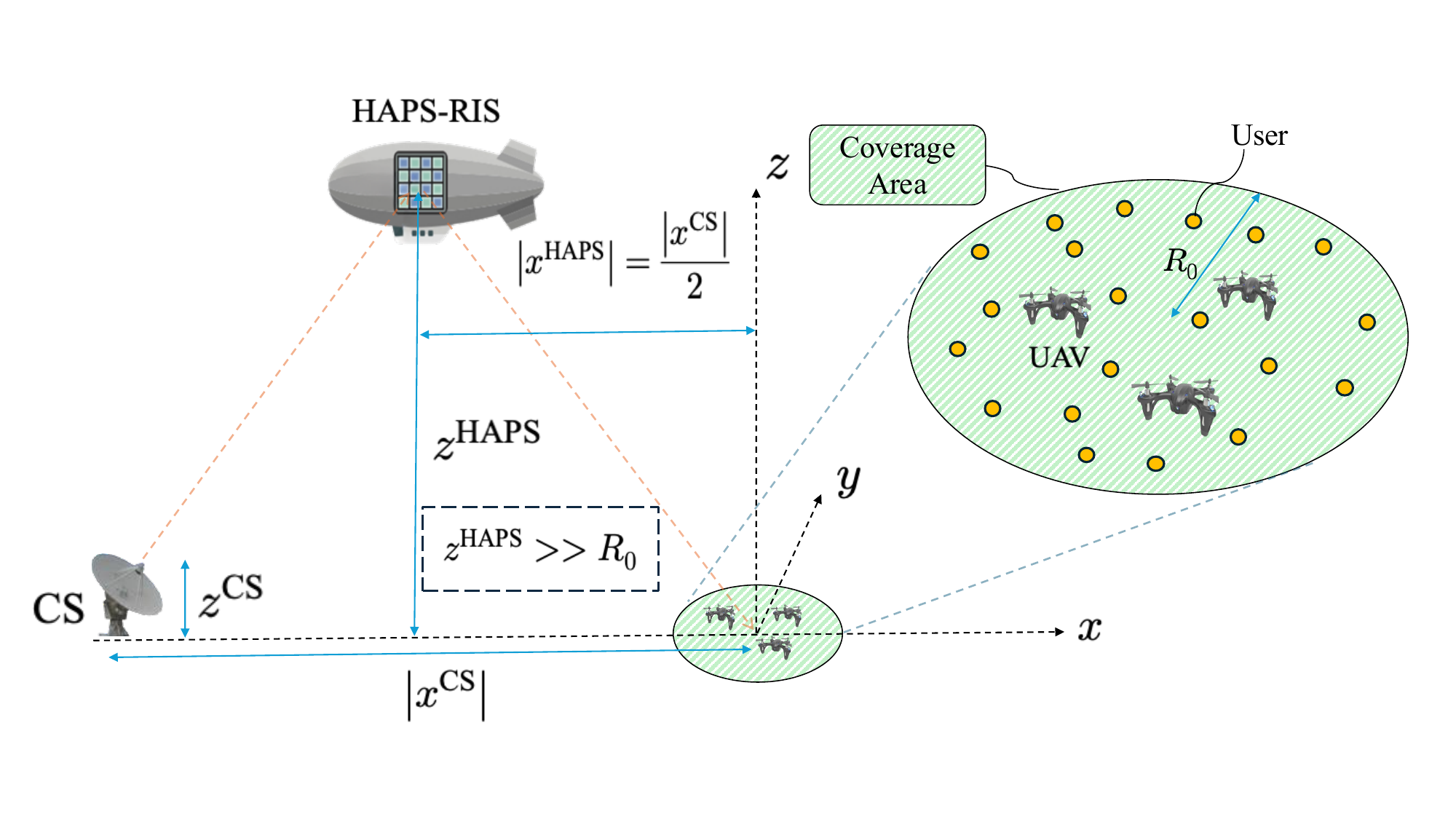}
    \caption{The Proposed System Model}
    \label{sys1}
\end{figure}

We consider a downlink transmission scenario where heterogeneous non-terrestrial infrastructures, i.e., HAPS--RIS and UAVs, are exploited to bring connectivity to a remote area as shown in Fig. \ref{sys1}. We consider that the UAVs and the CS operate independently to provide connectivity and they are not connected to each other. 
Severe blockage eliminates the possibility of direct CS–user links, making the HAPS-RIS-assisted cascaded channel the only viable communication path.
The set of \textcolor{black}{stationary} users $\mathcal{I}$ is defined as $\mathcal{I}$=\{1, \dots, $I$\}, where $i \in \mathcal{I}$ represents an individual user indexed by $i$. The users are uniformly distributed in a circular area, with a minimum separation distance of $D_{0}$
among them, where there is no nearby terrestrial infrastructure. 

The set of UAVs $\mathcal{J}$ is defined as $\mathcal{J}$ = \{1, \dots, $J$\}, where $j \in \mathcal{J}$ represents an individual UAV indexed by $j$.
\textcolor{black}{In this scenario, the UAVs serve as hovering aerial base stations. In HAPS-RIS scenarios, CS uses high gain directional antenna, creating a dominated LoS connection with HAPS \cite{alfattani2021aerial,alfattani2021link, alfattani2022beyond, alfattani2023resource, tekbiyik2021channel,azizi2023ris}.}

The set of RIS elements $\mathcal{M}$ is defined as $\mathcal{M}$ = \{1, \dots, $M$\}, where $m \in \mathcal{M}$ represents the RIS element indexed by $m$.
 Users can be connected to the network either through the UAV system or the HAPS-RIS setup, but not through both at the same time. The coverage area is a circle with the radius $R_0$ with the centre of $(x_0,y_0, z_0)$, the CS location is $(x^{\rm{CS}},y^{\rm{CS}},z^{\rm{CS}})$, the location of the HAPS is $(x^{\rm{HAPS} },y^{\rm{HAPS} },z^{\rm{HAPS}})$, and the location of the UAV $j$ is $(x_{j}^{\rm{UAV} },y_{j}^{\rm{UAV} },z_{j}^{\rm{UAV}})$ in the Cartesian coordinate system. 
 The set of equally spaced orthogonal subcarriers $\mathscr{L}$ is defined as $\mathscr{L}$=\{1, \dots, $L^{\rm{tot}}$\}, where $l \in \mathscr{L}$ represents the subcarrier indexed by $l$. The system bandwidth $BW$ is divided into $L^{\rm tot}$ equally spaced and orthogonal subcarriers to be shared between the CS and the UAVs for serving the users. Therefore, the bandwidth allocated to each orthogonal subcarrier is $B_{l}=\frac{BW}{L^{\rm{tot}}}$. 

\subsection{UAV-based Network Architecture}
 We consider a cellular scenario where the users covered by multiple UAVs are using orthogonal frequency division multiple access (OFDMA), and hence, there are multi-cells where there is inter-cell interference but no intra-cell interference.
The signal to interference plus noise ratio (SINR) of the user $i$ served by UAV $j$ over subcarrier $l'$ can be obtained as 
\begin{equation}
\begin{array}{l}
        \gamma^{l',\rm{UAV} }_{ij}=
        \frac{P^{l',\rm{UAV} }_{ij}\tilde{\rho}^{l',\rm{UAV} }_{ij} \left| h^{\rm{UAV} }_{ij} \right|^{2}  }{N_{0}B_{l'}+\sum^{J}_{j^{\prime }\neq j} P^{l',\rm{UAV} }_{ij^{\prime }}\tilde{\rho}^{l',\rm{UAV} }_{ij^{\prime }} \left| h^{\rm{UAV} }_{ij^{\prime }} \right|^{2}  }, \forall i, \forall j, l'
\end{array}
\end{equation} 
where $N_{0}$ is the noise power spectral density and $P^{l',\rm{UAV} }_{ij}$ is the transmit power of the UAV $j$ allocated to the subcarrier $l'$ and the user $i$. $\tilde\rho^{l',\rm{UAV}}_{ij}$ is a binary variable where $\tilde\rho^{l',\rm{UAV}}_{ij}=1$ indicates that subcarrier $l'$ of UAV $j$ is assigned to user $i$, while $\tilde\rho^{l',\rm{UAV}}_{ij}=0$ otherwise. $h^{\rm{UAV} }_{ij}$ is the channel gain between user $i$ and UAV $j$, which can be formulated as $    h^{\rm{UAV} }_{ij}=\sqrt{G^{\rm{UAV} }_{j}G_{i}\bar{\mathcal{L}}^{-1}_{ij}  }, \forall i, j$. $G^{\rm{UAV} }_{j}$ is the antenna gain of the UAV $j$, $G_{i}$ is the antenna gain of the user $i$, and $\bar{\mathcal{L}}_{ij}$ is the average path loss between user $i$ and UAV $j$ which can be obtained as \cite{al2014optimal}
\begin{equation}\label{average pathloss uav}
\hspace{-1mm}
    \begin{array}{l}
          \bar{\mathcal{L}}_{ij}= P^{\rm{UAV} }_{ij,\rm{LoS} }\eta_{1} \left( \frac{4\pi f_{\rm c}}{c} d_{ij}\right)^{\alpha }+ P^{\rm{UAV} }_{ij, \rm{NLoS}}\eta_{2} \left( \frac{4\pi f_{\rm c}}{c} d_{ij}\right)^{\alpha }, \forall i, j,  
    \end{array}
\end{equation}
where $f_{\rm c}$ is the carrier frequency and $c$ is the speed of wave propagation which is considered to be equal to the speed of light. $\alpha$ is the path loss exponent, and $\eta_{1}$ and $\eta_{2}$ are the excessive path loss 
coefficients for LoS and non-LoS (NLoS) connections, respectively. 
The distance between user $i$ and UAV $j$ can be formulated as 
\begin{equation}
    d_{ij}=\sqrt{\left( x_{i}-x^{\rm{} \rm{UAV} }_{j}\right)^{2}  +\left( y_{i}-y^{\rm{UAV} }_{j}\right)^{2}  +(z_{i}-z^{\rm{UAV} }_{j})^{2}}, \forall i, j.
\end{equation}
 $P^{\rm{UAV} }_{ij,\rm{} \rm{LoS} }$ is the LoS probability between the user $i$ and the UAV $j$ as $  P^{\rm{UAV} }_{ij,\rm{} \rm{LoS} }=\frac{1}{1+\psi e^{-\beta \left( \tilde{\theta}_{ij} -\psi \right)  }}, \forall i, j,$
where $\beta$ and $\psi$ are the constant values that depend on the environment type, $  P^{\rm{UAV} }_{ij,\rm{} \rm{LoS} }+P^{\rm{UAV} }_{ij,\rm{} \rm{NLoS} }=1, \forall i, \forall j,
$ and $ \tilde{\theta}_{ij} =\frac{180}{\pi } \arcsin \left( \frac{z^{\rm{UAV} }_{j}}{d_{ij}} \right),  \forall i, \forall j$.

\subsection{HAPS--RIS assisted Network Architecture}
In the HAPS--RIS assisted network, CS communicates with the users through HAPS--RIS.
The coordinates of the CS, user $i$, HAPS, and RIS element $m$, are $(x^{\rm{CS}},y^{\rm{CS} },z^{\rm{CS} })$, 
 $\left( x_{i},y_{i},z_{i}\right)$,
 $(x^{\rm{HAPS} },y^{\rm{HAPS} },z^{\rm{HAPS}})$, $\left( x^{\rm{RIS} }_{m},y^{\rm{RIS} }_{m},z^{\rm{RIS} }_{m}\right)$, respectively. To define the channel, we consider $h^{\rm{CS} }_{im}$ as the cascade channel gain between user $i$-RIS and RIS-CS through RIS element $m$.
 
\begin{equation}
    h^{\rm{CS} }_{im}=\sqrt{G^{\rm{CS} }G_{i}\left( \mathcal{L}^{\rm{CS-RIS} }_{m}\right)^{-1}  \left( \mathcal{L}^{\rm{RIS-user} }_{im}\right)^{-1}},  \forall m,i,
\end{equation}
where $G^{\rm{CS}}$ is the antenna gain of the CS and $G_{i}$ is the antenna gain of the user $i$. Using the Friis model, \cite{friis1946note}, the path loss between CS and RIS element $m$ can be formulated as 
\begin{equation}
    \begin{array}{l}
         \mathcal{L}^{\rm{CS}-\rm{RIS} }_{m}=(\frac{4\pi f_{\rm c}}{c} )^{2}\times\\
         \left( \left( x^{\rm{RIS} }_{m}-x^{\rm{CS} })^{2}+(y^{\rm{RIS} }_{m}-y^{\rm{CS} })^{2}+(z^{\rm{RIS} }_{m}-z^{\rm{CS}}\right)^{2}  \right)  ,\  \forall m,
    \end{array}
\end{equation}
and the path loss between RIS element $m$ and the user $i$ can be formulated as
\begin{equation}
    \begin{array}{l}
         \mathcal{L}^{\rm{RIS}, \rm{user}}_{im} = \left( \frac{4\pi f_{\rm{c}}}{c} \right)^{2} \times \\
 \left( (x^{\rm{RIS}}_{m} - x_{i})^{2} + (y^{\rm{RIS}}_{m} - y_{i})^{2} + (z^{\rm{RIS}}_{m} - z_{i})^{2} \right), \forall m, i,
    \end{array}
\end{equation}
The beyond cell users, which are covered by HAPS-RIS, are orthogonal to each other and to the cellular users served by the UAVs. Hence, there is no inter-user interference. The signal to noise ratio (SNR) of the received signal at user $i$ and subcarrier $l$, can be obtained as 
\begin{equation}\label{SNRHAPS}
   \gamma^{l,\rm{CS} }_{i}=\frac{{P_{i}^{l,\rm{CS} }}\left|\sum \limits_{m=1}^{M}\rho^{l,\rm{CS} }_{im} h^{\rm{CS} }_{im}\vartheta_{im} \right|^{2}  }{N_{0}B_{l}}, \forall i, l,
 \end{equation}
 where $M$ is the total number of RIS elements and $P_{i}^{l,\rm{CS}}$ denotes the transmit power allocated by the CS to the user $i$ over the subcarrier $l$. $\rho^{l,\rm{CS} }_{im}$ is the binary variable which demonstrates the RIS element $m$ is assigned to the user $i$ over subcarrier $l$.
 Considering $\vartheta_{im}$ as the reflection coefficient of the RIS element $m$ corresponding to the user $i$ as 
\begin{equation}
\vartheta_{im} =\mu_{m} e^{-j(\phi_{m} -\xi^{\rm{CS} }_{m} -\omega_{im} )}, \forall i,m,
\end{equation}
where $\phi_{m}\in [0,2\pi]$ is the RIS phase shift for the element $m$. $\xi^{\rm{CS} }_{m}$ is the corresponding phase between the RIS element $m$ and the CS. $\omega_{im}$ is the corresponding phase between RIS element $m$ and the user $i$. $0 \le \mu_m \le 1$
is the reflection loss corresponding to the RIS element $m$. 
\\
 

\section{The Proposed Multi-objective Joint Optimization Problem}
To fully integrate the HAPS-RIS and UAV-based network architectures, we propose a multi-objective joint optimization problem that simultaneously considers the key parameters of both infrastructures. The objective functions include: (i) the number of users covered by the HAPS–RIS system, (ii) the required number of UAVs and (iii) the total average air-to-ground path loss experienced by users served by UAVs. These objectives reflect the dual nature of the integrated system, where the first objective corresponds to the HAPS–RIS setup and the latter two are associated with the UAV network.

\vspace{2mm}
\textbf{\textit{Definition 1}}: The total number of users covered by the HAPS–RIS is obtained by counting the number of users that are assigned to at least one RIS element on at least one subcarrier. User $i$ is considered covered by the HAPS–RIS if there exists \emph{any} RIS element $m$ and \emph{any} subcarrier $l$ such that $\rho^{l,\rm{CS}}_{im}=1$.
While $\rho^{l,\rm{CS}}_{im}=0$ means that no transmission is scheduled for this user–RIS–subcarrier combination. Based on this rule, the total number of HAPS–RIS–served users can be obtained as
\begin{equation}\label{firstobjF}
    U^{\text{HAPS} }=\sum^{}_{i \in \mathcal{I}} \max_{\forall m,l} \rho^{l, \rm{CS}}_{im},
\end{equation}
where the inner maximization checks whether user $i$ is assigned to at least one RIS element and one subcarrier.

\vspace{2mm}
\textbf{\textit{Definition 2}}: The total average path loss between UAVs and ground users is obtained by aggregating the path loss values over all UAV–user pairs that are actually assigned to at least one subcarrier. The total average UAV path loss over all associated UAV–user pairs is obtained as
\begin{equation}\label{third obj}
\varLambda^{\text{UAV} } =\sum^{}_{i \in \mathcal{I}} \sum^{}_{j \in \mathcal{J}} (\max_{\forall l'} \tilde\rho^{l', \rm{UAV}}_{ij} )\bar{\mathcal{L} }_{ij},
\end{equation}
where the inner maximization term acts as an association indicator; it is equal to one if user $i$ is served by UAV $j$ on at least one subcarrier, and zero if no subcarrier of UAV $j$ is allocated to user $i$. $\bar{\mathcal{L}}_{ij}$ denotes the average  path loss between UAV $j$ and user $i$, obtained in \eqref{average pathloss uav}.  

In the proposed multi-objective joint optimization framework, we aim to maximize the number of covered users by the HAPS–RIS, minimize the number of required UAVs, and minimize the overall average UAV path loss. The optimization variables associated with the UAV setup include the two-dimensional placement of the UAVs, i.e., their latitude and longitude, the binary association indicators for the UAV–user–subcarrier assignments, and the number of UAVs required to serve the users. The optimization variables associated with the HAPS–RIS setup include the binary association indicators for the HAPS–RIS subcarrier–user–RIS element assignments and the RIS phase-shifts. The constraints ensure that all users satisfy their quality-of-service (QoS) requirements, that the power budget limitations at both the CS and the UAVs are met, and that the OFDMA-based orthogonality and user-association rules are guaranteed.
The main optimization problem can be designed as
\begin{subequations}
\begin{equation}\label{4 obj func}
\begin{aligned}
\textbf{OP 1:}  
\max_{\boldsymbol{X}^{\mathrm{UAV}},
        \boldsymbol{Y}^{\mathrm{UAV}},
        \boldsymbol{\rho}^{\mathrm{CS}},
        \tilde{\boldsymbol{\rho}}^{\mathrm{UAV}},
        N^{\mathrm{UAV}},
        \boldsymbol{\phi}}
\bigl[ U^{\mathrm{HAPS}},\,-N^{\mathrm{UAV}},\,\\-\varLambda^{\mathrm{UAV}} \bigr]
\end{aligned}
\end{equation}
\begin{equation}\label{binary values cons}
\hspace{-2cm}
   {{\rm{s} }  .{\rm{t} }  .\  }   \tilde\rho^{{l'},\rm{UAV} }_{ij} ,\rho^{l,\rm{CS} }_{im} \in \{ 0,1\} \ \  \forall i,j,m,l,l',
\end{equation}
\begin{equation}\label{user min rate cons}
 \hspace{0.7cm}
\begin{array}{l}
 \sum\limits^{N^{\text{UAV} }}_{j=1} \sum\limits^{L^{\text{tot} }}_{l'=1} \log_{2} \left( 1+\gamma^{l',\rm{UAV} }_{ij} \right)  +\sum\limits^{L^{\text{tot} }}_{l=1} \log_{2} \left( 1+\gamma^{l,\rm{CS} }_{i} \right) \\ \geq r^{\rm{user} }_{0} \  \forall i\in\mathcal{I},
\end{array}
\end{equation}
\begin{equation}\label{user association cons}
\hspace{-1.7cm}
\rho^{l,\rm{CS} }_{im} +\sum^{J}_{j=1} \tilde\rho^{l',\rm{UAV}}_{ij} \leq 1\  \  \forall i,l,l^{\prime },m,
\end{equation}
\begin{equation}\label{max power CS cons}
\hspace{-2.2cm}
    \sum^{I}_{i=1} \sum^{M}_{m=1} \sum^{L^{\text{tot} }}_{l=1} P^{l,\rm{CS} }_{i}\rho^{l,\rm{CS} }_{im} \leq P^{\rm{CS} }_{\rm{max} },
\end{equation}
\begin{equation}\label{max power uav cons}
\hspace{-1.7cm}
\sum^{I}_{i=1} \sum^{L^{\text{tot} }}_{l'=1} P^{l',\rm{UAV} }_{ij}\tilde\rho^{l',\rm{UAV} }_{ij} \leq P^{j}_{\rm{max} } ~~\forall j,
\end{equation}
\begin{equation}\label{RIS unit assignment cons}
\hspace{-2.2cm}
    \rho^{l,\rm{CS} }_{im} +\rho^{\tilde{l},\rm{} \rm{CS} }_{i^{\prime }m} \leq 1\  \  \forall i\neq i^{\prime },l,\tilde{l},m,
\end{equation}
\begin{equation}\label{Subcarrier orthogonality cons}
\hspace{-1.9cm}
 \sum^{I}_{i=1} (\rho^{l,\rm{CS} }_{im} +\tilde\rho^{l',\rm{} \rm{UAV} }_{ij} )\leq 1\  \forall l,l',j,m,
\end{equation}
\end{subequations}
where $X^{\text{UAV} }=\left[ x^{\text{UAV} }_{1},...,x^{\text{UAV} }_{N^{\text{UAV} }}\right]_{1\times N^{\text{UAV} }} $ and $  Y^{\text{UAV} }=\left[ y^{\text{UAV} }_{1},...,y^{\text{UAV} }_{N^{\text{UAV} }}\right]_{1\times N^{\text{UAV} }} $ are the vectors representing the latitudes and longitudes of the UAVs, respectively.
Constraint \eqref{binary values cons} demonstrates the binary variables. Constraint \eqref{user min rate cons} ensures that the data rate received by each user meets the minimum data rate requirement where the minimium threshold is $r^{\rm{user} }_{0}$ for all the users. Although \eqref{user min rate cons} includes both the HAPS–RIS and UAV rate terms, a user cannot be served by both simultaneously. The inequality \eqref{user association cons} guarantees that each user can be connected to either HAPS-RIS setup or a UAV, but not both of them. Thus, each user is linked either to the HAPS–RIS or to a single UAV, and the other rate term in \eqref{user min rate cons} becomes zero accordingly. Furthermore, \eqref{user association cons} ensures that each user can not be connected to more than one UAV.
In \eqref{max power CS cons}, we ensure that the total transmission power allocated to users covered by the HAPS-RIS setup does not exceed $P^{\rm{CS} }_{\rm{max}}$ which is the maximum power limit of the CS. In \eqref{max power uav cons}, we ensure that the total transmission power allocated to the users supported by the UAV $j$ does not exceed $P^{{j}}_{\rm{max}}$ which is the maximum power limit of the UAV $j$. 
Constraint \eqref{RIS unit assignment cons} guarantees that each RIS element can not be assigned to more than one user. Given with the assumption that the single CS uses OFDMA, all the users covered by HAPS-RIS setup are considered as a single cell, so that there is no inter-user interference. Hence, a dedicated set of RIS units is allocated to each user on its assigned subcarriers. This is because RIS elements can operate differently across frequencies \cite{bjornson2022reconfigurable}. Constraint \eqref{Subcarrier orthogonality cons} guarantees that the subcarriers allocated to each of the users supported by HAPS-RIS are orthogonal to the subcarriers associated to other users. In addition, this constraint ensures that the subcarriers allocated to the users supported by HAPS-RIS are orthogonal to the subcarriers allocated to the UAVs. This assumption is in line with the assumptions in HAPS-RIS scenerios, \cite{alfattani2022beyond,azizi2025exploring}, where the CS is considered as the auxiliary infrastructure providing a beyond cell over the existing network. Furthermore, this constraint guarantees that the allocated subcarriers to each UAV are orthogonal to other UAVs' assigned subcarrier. This is because we consider UAVs use OFDMA so that there is no intra-cell interference but there is inter-cell interference.

\section{The Proposed Solution Method}
The proposed optimization problem \textbf{OP~1} is a large-scale mixed-integer nonlinear program (MINLP) formulated as a multi-objective joint design over tightly coupled variables. These variables include the two-dimensional placement of UAVs, the required number of UAVs, the RIS phase shifts, and the user association indicators across UAVs, RIS elements, and subcarriers. The heterogeneous nature of the integrated HAPS--RIS and UAV architecture, combined with the massive number of RIS elements, users, and subcarriers, renders \textbf{OP~1} a high-dimensional and highly non-convex MINLP. Such formulations are well known to be NP-hard, making the computation of an exact optimal solution intractable.
The intrinsic complexity of \textbf{OP~1} stems from three tightly coupled components. The UAV placement problem introduces a non-convex continuous optimization challenge, as the UAV channel gains are functions of the UAV locations and are jointly coupled with communication performance. The RIS phase-shift design further increases the dimensionality of the problem due to the need to jointly optimize a massive number of reflecting elements. In addition, the user association decisions give rise to three-dimensional binary assignment structures, namely user--subcarrier--RIS element and user--subcarrier--UAV associations, resulting in a high dimensional combinatorial optimization. To overcome this intractability, our goal in this section is to provide a sequence of transformations included in \textit{Theorem 1}, \textit{Lemma 1}, and \textit{Proposition 1}, which relax and map \textbf{OP~1} into a tractable optimization problem. This principled sequence of analytical transformations provides a clear pathway from the original NP-hard formulation to a low-complexity solution framework.
\subsection{Proposed Upper Bound and $k$-Means Equivalence}


 \textcolor{black}{\textbf{\textit{Theorem 1}}: \textbf{OP 1} can be relaxed into \textbf{OP 2}  with two objective functions as}
\begin{subequations}
    \begin{equation}
    \hspace{-1.5cm}
\textcolor{black}{\textbf{OP 2}:~~~ \max_{\boldsymbol{\rho}^{\rm{CS} } ,\boldsymbol{\tilde\rho}^{\rm{UAV} } ,\  N^{\rm{UAV} },\boldsymbol{\phi }}} \  [U^{\rm{HAPS} },  -N^{\rm{UAV} }]
\end{equation}
\begin{equation}\label{ratecons2}
    \hspace{-0.7cm}
\textcolor{black}{ {{\rm{s}}.{\rm{t}}.:} ~~ \eqref{binary values cons}-\eqref{Subcarrier orthogonality cons}}  
\end{equation}
\end{subequations} 
where $X^{\rm{UAV}}$, $Y^{\rm{UAV}}$ and $\zeta_{ij} =\max\limits_{\forall l'}{\tilde\rho^{l',\rm{UAV}}_{ij}}$ can be obtained based on $k$-means clustering operator, with the $k$ factor equal to $N^{\rm{UAV}}$. Since the upper bound of \eqref{third obj} is minimized through the $k$-means clustering operator, the third objective function of \textbf{OP 1} no longer needs to be included. This Theorem is valid for $\alpha=2$ and the fixed UAV altitudes.

\begin{proof} 
In order to relax the third objective function, the aim is to obtain an upper bound for $\varLambda^{\text{UAV} }$. By substituting \eqref{average pathloss uav} into \eqref{third obj}, we have
    \begin{equation}\label{third obj equival}
        \varLambda^{\text{UAV}}=\sum_{i} \sum^{}_{j} \zeta_{ij}\left( P^{\rm{UAV} }_{ij,\rm{LoS} }\eta_{1} +P^{\rm{UAV} }_{ij,\rm{NLoS} }\eta_{2} \right)  \left( \frac{4\pi f_{\rm c}}{c} d_{ij}\right)^{\alpha }.
    \end{equation}
    Due to the fact that $1\geq P^{\rm{UAV} }_{ij,\rm{LoS} },P^{\rm{UAV} }_{ij,\rm{NLoS} }\geq0,\  \forall i,j,$ and $\eta_2>\eta_1$ we have 
        \begin{equation}\label{plos neq1}
    \begin{array}{l}
       P^{\rm{UAV} }_{ij,\rm{LoS} }\eta_{1} +P^{\rm{UAV} }_{ij,\rm{NLoS} }\eta_{2}= \\\\P^{\rm{UAV} }_{ij,\rm{LoS} }\eta_{1} +\left( 1-P^{\rm{UAV} }_{ij,\rm{LoS} }\right)  \eta_{2} =P^{\rm{UAV} }_{ij,\rm{LoS} }\left( \eta_{1} -\eta_{2} \right)  +\eta_{2} \leq \eta_2,
    \end{array}
    \end{equation}
    By substituting \eqref{plos neq1} into \eqref{third obj equival},
it can be observed that for $\alpha=2$, the fixed UAV altitudes $z_{j}=z_{0, \text{UAV}}, \forall j$, without loss of generality by considering $z_{i}=0, \forall i$, and given the fact that $0\leq\zeta_{ij} \leq 1, \forall i,j,$ and $N^{\rm{UAV}}\leq N^{\rm{UAV},(0)}$, the upper bound of \eqref{third obj equival} can be obtained as
\begin{equation}\label{third obj upp}
\begin{aligned}
\varLambda^{\text{UAV,upp}}
&= \eta_{2} \left( \frac{4\pi f_{\rm c}}{c} \right)^{2} \Bigg(
z_{0,\text{UAV}}^{2} I N^{\rm UAV,(0)}  \\
&\quad + \sum_{i} \sum_{j}^{N^{\text{UAV}}} \zeta_{ij}
\left( (x_i-x_j)^2 + (y_i-y_j)^2 \right)
\Bigg),
\end{aligned}
\end{equation}
Interestingly, it can be observed that minimizing \eqref{third obj upp} is equivalent to run the $k$-means clustering as 
\begin{equation}
\begin{array}{l}
    \min \varLambda^{\text{UAV,upp} } \triangleq\eta_{2} \left( \frac{4\pi f_{\rm c}}{c} \right)^{2}\times\\  \left( \underbrace{\min_{\forall x_{j}\in X,\forall y_{j}\in Y,\zeta_{ij} } \sum_{i} \sum^{N^{\text{UAV} }}_{j} \zeta_{ij} \left( \left( x_{i}-x_{j}\right)^{2}  +\left( y_{i}-y_{j}\right)^{2}  \right)  }_{k-\text{means clustering} } \right),
\end{array}
\end{equation} 
where the $k$ factor is equal to $N^{\rm{UAV}}$ and cluster centers are UAV locations. Each user $i$ is assigned to its nearest cluster center, which determines its serving UAV. Hence, the cluster membership indicator plays the role of the UAV--user association variable, i.e., $\zeta_{ij} = 1$ if user $i$ belongs to the cluster associated with UAV $j$, and $\zeta_{ij} = 0$ otherwise. This mapping reveals that minimizing the sum of squared distances in $k$-means is equivalent to minimizing the upper bound of the total average UAV path loss in \eqref{third obj}. Therefore, once $k$-means is applied for a given $N_{\text{UAV}}$, the UAV latitude and longitudes and UAV–user associations become deterministic outputs rather than optimization variables. As the upper bound of the \eqref{third obj} is minimized via $k$-means clustering, \textbf{OP 1} can be relaxed into \textbf{OP 2}.
\end{proof}
\subsection{Proposed Practical RIS Phase Shift Design}
Although \textbf{OP~2} already reduces the complexity of the main optimization problem, it remains a large-scale non-convex MINLP, and hence, NP-hard. Owing to the fact that there are massive number of RIS elements in the HAPS-RIS scenarios, obtaining the optimal RIS phase shifts in this high complexity problem leads to huge computational overhead which is intractable. Hence, the target in this step is to define a practical RIS phase shift design to relax \textbf{OP 2} into a problem with a lower computational complexity.

\vspace{2mm}
 \textcolor{black}{\textbf{\textit{Lemma 1}}: \textbf{OP 2} can be relaxed into }
\begin{subequations}
\begin{equation}
   \hspace{-2cm}
\textcolor{black}{\textbf{OP 3}:~~~ \max_{\boldsymbol{\rho}^{\rm{CS} } ,\boldsymbol{\tilde\rho}^{\rm{UAV} } ,\  N^{\rm{UAV}}}} \  [U^{\rm{HAPS} },  -N^{\rm{UAV} }]
\end{equation}
\begin{equation}\label{ratecons3}
    \hspace{-1.3cm}
\textcolor{black}{ {{\rm{s}}.{\rm{t}}.:} ~~ \eqref{binary values cons}-\eqref{Subcarrier orthogonality cons}}
\end{equation}
\end{subequations}
\textcolor{black}{with the proposed practical RIS phase shift design $\phi^{\ast }_{m}= \xi_{0} +\omega_{0} +2\tilde{k}\pi, ~\forall m$ where $\tilde{k}$ is an integer value.}
\begin{proof} 
In the proposed geometrical case, the dimensions of RIS can be negligible compared to the altitude of HAPS, as it does not have much effect on the performance. Hence, without loss of generality, we relax the location of the RIS elements as $\begin{matrix}\left( x^{\rm{RIS} }_{m},y^{\rm{RIS} }_{m},z^{\rm{RIS} }_{m}\right)  \approx (x^{\rm{HAPS} },y^{\rm{HAPS} },z^{\rm{HAPS} }),&\forall m\end{matrix}.$
\footnote{\textcolor{black}{The dimensions of the RIS are in the range of 100 meters while the HAPS-to-CS and HAPS-to-users distances are in the range of more than 20 kilometers.}}
The dimensions of the RIS and the coverage area is much less than the range of the distance between HAPS-RIS and the terrestrial components. Accordingly, without loss of generality, we can relax the corresponding phases to fixed values as $\xi^{\rm{CS} }_{m} \approx \xi_{0}, \forall m $ and $\omega_{im} \approx \omega_{0}, \forall i\in\mathcal{C}, \forall m,$ so the reflection gain of RIS element $m$ can be relaxed as $\vartheta_{im} \approx {\tilde\vartheta_{m}}, \forall i\in\mathcal{C}, \forall m$ where
\begin{equation}\label{relaxedreflection}
{\tilde\vartheta_{m}}= \mu_{m} e^{-j(\phi_{m} -\xi_{0} -\omega_{0} )}\  \  \  \forall m.
\end{equation}
\textcolor{black}{As \eqref{relaxedreflection} is independent from user index, $r^{\rm{CS} }_{i}(R,\tilde\vartheta_{m})$ can be relaxed into its upper bound by considering $\phi^{\ast }_{m} =\arg \max {\tilde\vartheta_{m} } =\xi_{0} +\omega_{0} +2\tilde{k}\pi, \forall m$. Accordingly, \eqref{ratecons2} is relaxed into \eqref{ratecons3} and hence \textbf{OP 2} is relaxed into \textbf{OP 3}. This provides a practical implementation approach, given the massive number of RIS elements in HAPS-RIS scenario.}
\end{proof}

\subsection{Proposed Mapping Technique}
Although \textbf{OP~3} significantly reduces the dimensionality of the original formulation, it remains a large-scale non-convex MINLP and is therefore still NP-hard. The remaining complexity arises from the combinatorial user-UAV-subcarrier and user-subcarrier-RIS-element association variables. Jointly optimizing these discrete assignments is computationally prohibitive. To obtain a tractable problem while preserving the essential coupling between the HAPS--RIS and UAV infrastructures, we introduce a new set of geometric and bandwidth-related optimization variables and map the combinatorial associations in \textbf{OP~3} into these lower-dimensional variables. This leads to the proposed optimization problem \textbf{OP~4}, stated below.

\vspace{5mm}
\textcolor{black}{\textbf{\textit{Proposition 1}}: \textbf{OP 3} can be \emph{relaxed} and \emph{mapped} into \textbf{OP 4} as}
\begin{subequations}
    \begin{equation}\label{3 obj func}
    \hspace{-2.5cm}
\textcolor{black}{\textbf{OP 4}:~~~\max_{ R, N^{\rm{UAV} }, \varkappa}} \  [\tilde{U}^{\rm{HAPS} },  -N^{\rm{UAV} } ]
\end{equation}
\begin{equation}\label{binary values consop4}
\hspace{-3cm}
{{\rm{s} }  .{\rm{t} }  .\  } \tilde{\rho}^{{l^{\rm{u}}},\rm{UAV} }_{ij} \in \{ 0,1\} \  ,\forall i\in \mathcal{B},j,l^{\text{u} },\  
\end{equation}
\begin{equation}\label{binary values 2 consop4}
\hspace{-4cm}
\tilde{\rho}^{{l^{\rm{c}}},\rm{CS} }_{im} \in \{ 0,1\} ,\forall i\in \mathcal{C},m,l^{\text{c} },
\end{equation}
\begin{equation}\label{minrateUAV}
\hspace{1mm}
\sum\limits^{N^{\text{UAV} }}_{j=1} \sum\limits^{L^{\text{UAV} }}_{l^{\text{u}}=1} \log_{2} \left( 1+\tilde{\gamma}^{l^{\rm{u} },\rm{UAV} }_{ij} \left( R,\  N^{{}^{\text{UAV} }},\varkappa \right)  \right)  \geq r^{\rm{user} }_{0} ,\  \forall i\in \mathcal{B},
\end{equation}
\begin{equation}\label{minrateHAPS}
\hspace{-1.7cm} 
\sum\limits^{L^{{\text{CS} }  }}_{l^{\rm{c} }=1} \log_{2} \left( 1+\tilde{\gamma }^{l^{\rm{c} },\rm{CS} }_{i} (R,\varkappa )\right)  \geq r^{\rm{user} }_{0} \  \forall i\in \mathcal{C},
\end{equation}
\begin{equation}\label{user association consop4}
\hspace{-0.5cm}
\sum^{N^{\text{UAV} }}_{j=1} \tilde{\rho}^{\rm{UAV}, l^{\text{u} }}_{ij} \left( R,\  N^{\text{UAV} },\varkappa \right)  \leq 1\  \  \forall i\in \mathcal{B},l^{\text{u} },l^{\text{u}^{\prime } },
\end{equation}
\begin{equation}\label{max power CS consop4}
\hspace{-2.8cm}
\hspace{1cm} 
 \sum_{\forall i\in \mathcal{C}} \sum^{M}_{m=1} \sum^{L^{\text{CS} }}_{l^{\text{c} }=1} P^{l^{\rm{} \text{c} },\rm{CS} }_{i}\tilde{\rho}^{l^{\rm{} \text{c} },\rm{CS} }_{im} \left( R,\varkappa \right)  \leq P^{\rm{CS} }_{\rm{max} },\  
\end{equation}
\begin{equation}\label{max power uav consop4}
\hspace{2mm}
\sum_{\forall i\in \mathcal{B}} \sum^{L^{\text{UAV} }}_{l^{\text{u} }=1} P^{l^{\text{u} },\rm{UAV} }_{ij}\tilde{\rho}^{l^{\text{u} },\rm{UAV} }_{ij} \left( R,\  N^{\text{UAV} },\varkappa \right)  \leq P^{\rm{} j}_{\rm{max} }\forall j,\  
\end{equation}
\begin{equation}\label{RIS unit assignment consop4}
\hspace{3 mm}
\tilde{\rho }^{l^{\text{c} },\rm{CS} }_{im} \left( R,\varkappa \right)  +\tilde{\rho }^{l^{\text{c}^{\prime } },\rm{} \rm{CS} }_{i^{\prime }m} \left( R,\varkappa \right)  \leq 1\  \  \forall (i\neq i^{\prime })\in \mathcal{C},l^{\rm{} \text{c} },l^{\text{c} \prime },m,\end{equation}
\begin{equation}\label{Subcarrier orthogonality consop4}
\hspace{-3.8cm}
\sum^{}_{\forall i\in \mathcal{C}} \tilde{\rho }^{l^{\text{c} },\rm{CS} }_{im} \left( R,\varkappa \right)  \leq 1\  \forall l^{\rm{c} },m,\  \end{equation}
\begin{equation}\label{uav subcarrier orthogonal}
\hspace{-2.7cm}
\sum^{}_{\forall i\in \mathcal{B}} \tilde{\rho }^{l^{\text{u} },\rm{UAV} }_{ij} \left( R,N^{\text{UAV} },\varkappa \right)  \leq 1\  \forall l^{\rm{u} },j,\  
\end{equation}
\end{subequations}
where $R$ is the radius of the UAV zone, which is illustrated in the proposed geometric model in Fig. \ref{sys2}. $\varkappa$=$\frac{L^{\rm{CS}}}{L^{\rm{UAV}}}$ is the BW portioning factor where $L^{\rm{CS}}$=$\frac{\varkappa}{\varkappa +1}L^{\rm{tot}}$ and $L^{\rm{UAV}}$=$\frac{1}{\varkappa +1}L^{\rm{tot}}$ are the number of orthogonal subcarriers allocated to CS and UAVs, respectively, while $L^{\rm{tot}}$=$L^{\rm{CS}}$+$L^{\rm{UAV}}$. The combinatorial optimization variables of \textbf{OP 3} are relaxed and mapped into new optimization variables $R$ and $\varkappa$ and the constraints of \textbf{OP 3} in \eqref{binary values cons}-\eqref{Subcarrier orthogonality cons} are mapped as follows:
\eqref{binary values cons} $\rightarrow$ \eqref{binary values consop4}-\eqref{binary values 2 consop4};  
\eqref{user min rate cons} $\rightarrow$ \eqref{minrateUAV}-\eqref{minrateHAPS};  
\eqref{user association cons} $\rightarrow$ \eqref{user association consop4};  
\eqref{max power CS cons} $\rightarrow$ \eqref{max power CS consop4};  
\eqref{max power uav cons} $\rightarrow$ \eqref{max power uav consop4};  
\eqref{RIS unit assignment cons} $\rightarrow$ \eqref{RIS unit assignment consop4};  
\eqref{Subcarrier orthogonality cons} $\rightarrow$ \eqref{Subcarrier orthogonality consop4}–\eqref{uav subcarrier orthogonal}, and the objective function \eqref{firstobjF} is transformed into $\tilde{U}^{\rm{HAPS}}(R,\varkappa)$.

\begin{proof}
The key difficulty in \textbf{OP~3} lies in the three-dimensional binary association 
variables $\rho^{l,{\rm CS}}_{im}$ and $\tilde{\rho}^{l,{\rm UAV}}_{ij}$, which jointly
determine the user--RIS--subcarrier and user--UAV--subcarrier assignments, respectively. Since these
variables scale with the number of users, UAVs, RIS elements, and subcarriers,
their joint optimization results in an exponentially large combinatorial search space.
To construct a tractable reformulation, we exploit the geometric structure of the 
integrated HAPS--RIS and UAV system and map these high-dimensional variables into scalar variables.

\emph{1) Zone-based mapping of user associations:}
As illustrated in Fig.~\ref{sys2}, the coverage region is partitioned into two zones: an inner circular region of radius $R$, called \emph{UAV zone}, which is presented by the set $\mathcal{B} \subseteq \mathcal{I}$, and an outer ring between the main coverage and the UAV zone, called \emph{HAPS--RIS zone}, which is presented by the set $\mathcal{C} = \mathcal{I} \setminus \mathcal{B}$. All users in $\mathcal{C}$ are served by the HAPS--RIS and all users in 
$\mathcal{B}$ are served by the UAVs.
Due to the very large HAPS altitude relative to the ground coverage radius,
the channel gain between the HAPS--RIS and any ground user depends almost entirely 
on the HAPS altitude and is only marginally affected by the exact choice of user. Hence, the precise user association has negligible impact on 
performance. This allows us to replace the user-level association variables with the zone-level indicator.

\emph{2) Bandwidth portioning-based mapping of subcarrier allocations:} All UAVs are assumed to be homogeneous, operating at the same altitude with identical transmit powers, and all users have the same minimum rate requirement. The HAPS–RIS operates at a much higher altitude, and the CS has a different transmit-power budget than the UAVs. Moreover, users in the HAPS–RIS zone experience nearly symmetric channel conditions because the HAPS altitude is far greater than the coverage radius, making the impact of their exact locations negligible. Therefore, the detailed subcarrier assignment can be relaxed, and only the main heterogeneous factor distinguishing the two infrastructures, i.e., the bandwidth portioning ratio~$\varkappa$ between the HAPS–RIS and UAVs, is retained. The total bandwidth is partitioned into two orthogonal sub-bands corresponding to the UAV-served users and the HAPS–RIS-served users. This partitioning follows from the fact that the users of two zones are orthogonal to each other. Hence, the users do not interfere with one another. On this basis, we define two disjoint subcarrier sets $\mathscr{L}^{\rm UAV}=\{1,\dots,L^{\rm UAV}\}$, 
$\mathscr{L}^{\rm CS}=\{1,\dots,L^{\rm CS}\}$, where $\varkappa$=$\frac{L^{\rm{CS}}}{L^{\rm{UAV}}}$, $L^{\rm{tot}}$=$L^{\rm{CS}}$+$L^{\rm{UAV}}$, 
$L^{\rm UAV}=\frac{1}{\varkappa+1}L^{\rm tot}$ is the number of orthogonal subcarriers allocated to the UAV zone, and 
$L^{\rm CS}=\frac{\varkappa}{\varkappa+1}L^{\rm tot}$ is the number of subcarriers allocated to the HAPS–RIS zone. 
The two sets are mutually exclusive, i.e., $\mathscr{L}^{\rm UAV}\cap\mathscr{L}^{\rm CS}=\emptyset$ and $\mathscr{L}^{\rm UAV}\cup\mathscr{L}^{\rm CS}=\mathscr{L}^{\rm tot}$. This ensures that users in the HAPS–RIS zone and those in the UAV zone are served over orthogonal sub-bands without mutual interference. The bandwidth portioning variable $\varkappa$ thus encapsulates the resource coupling
between the two infrastructures, while eliminating the need for combinatorial
subcarrier allocations.

\emph{3) RIS clustering and association:}
Following the RIS-clustering strategy adopted in \cite{alfattani2023resource}, each user in the HAPS--RIS zone is assigned a dedicated cluster of RIS elements.
Given that the HAPS altitude is vastly larger than the maximum distance between RIS clusters, the choice of a specific RIS cluster has a negligible impact on the resulting channel gain. Consequently, the users in the HAPS–RIS zone are assigned to RIS clusters through a random injective mapping without loss of generality.

Combining the above, the only remaining optimization variables are the scalar-valued decision variables $R$, $N^{\rm UAV}$, and $\varkappa$, while all associations are deterministic functions of these variables. Hence, \textbf{OP~3} can be mapped into \textbf{OP~4}, completing the proof.
\qedhere
\end{proof}
\vspace{-0.5cm}
\subsection{Proposed Dynamic Pareto Optimization Framework} 
\begin{figure}
\hspace{-1.5mm}
\includegraphics[scale=0.27]{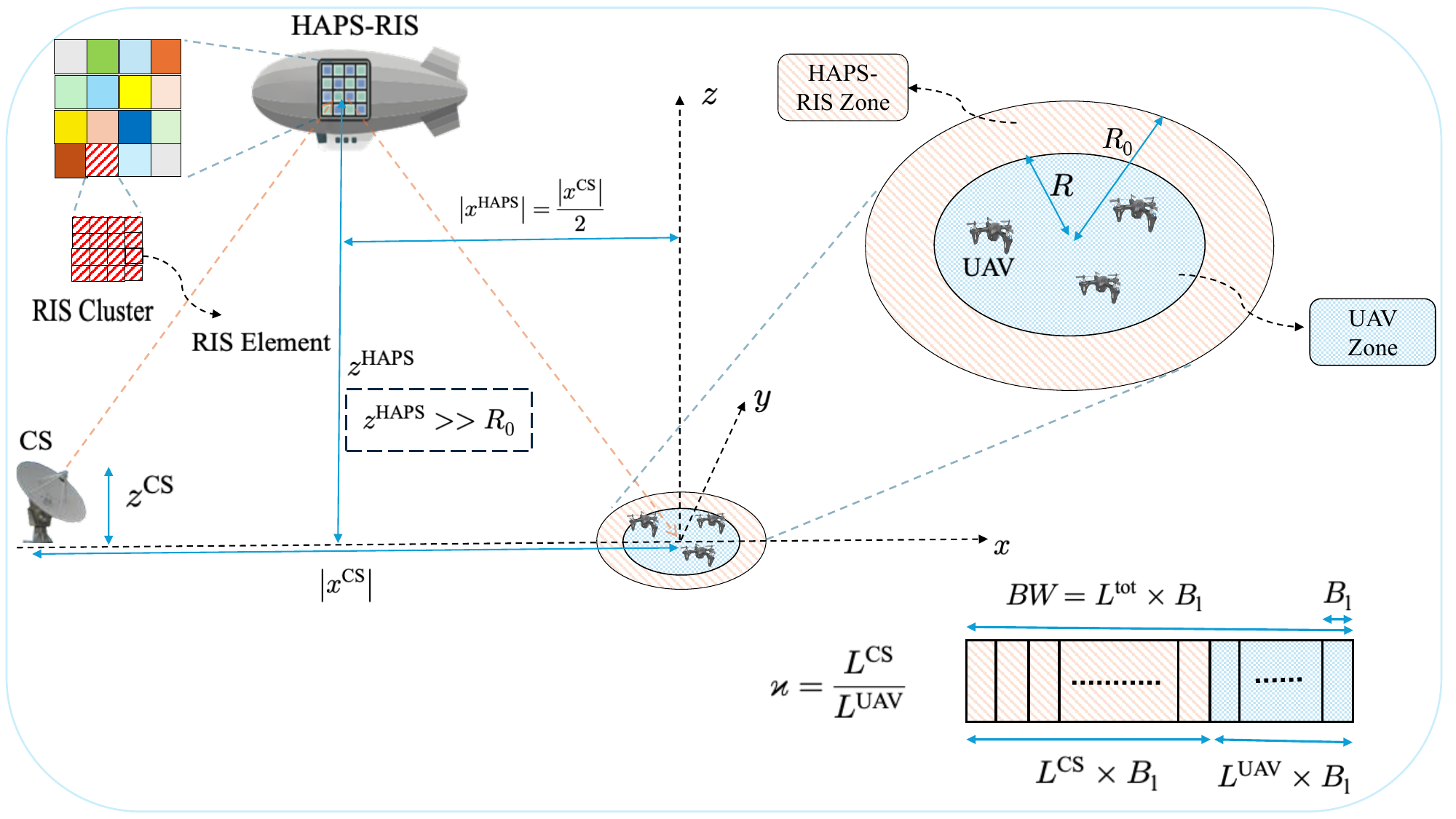}
    \caption{The proposed geometric model of \textbf{OP 4}}
    \label{sys2}
\end{figure}

\begin{figure}
\hspace{-9mm}
\includegraphics[scale=0.31]{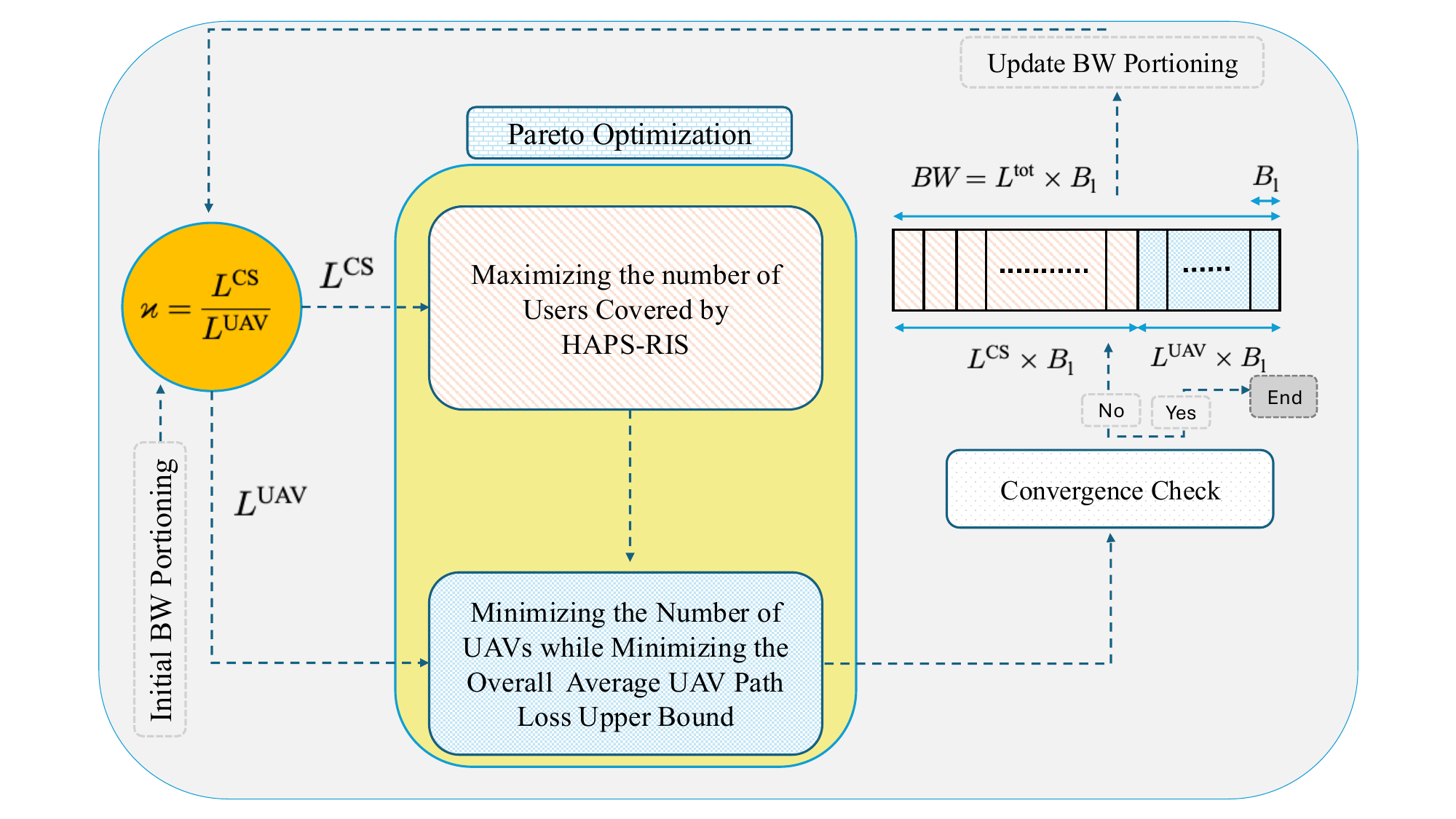}
    \caption{The proposed dynamic Pareto optimization method to solve \textbf{OP 4}} 
    \label{sys3}
\end{figure}

The proposed geometric model of \textbf{OP~4} is illustrated in Fig. \ref{sys2}, while the building blocks and the interfaces of the proposed dynamic Pareto optimization framework are shown in Fig. \ref{sys3}. \textbf{OP~4} is a joint multi-objective optimization problem in which the HAPS--RIS
and UAV systems are coupled through the bandwidth portioning factor.
This coupling variable determines the bandwidth portions which are allocated to the HAPS--RIS setup and UAV-based networks.
Therefore, the solution of OP~4 is structured around $\varkappa$ as the 
\emph{upper-layer} decision variable.  
For a given value of $\varkappa$, the objectives of OP~4 are
handled in the lower layer through a Pareto-based lexicographic optimizer \cite{miettinen1999nonlinear}. At each iteration, a candidate bandwidth portioning factor is selected and
passed to the Pareto optimization block in Fig.~\ref{sys3}. For this fixed $\varkappa$, the lexicographic Pareto process evaluates the 
three objectives.
We exploit pareto optimization, a hierarchical lexicographic approach.
The highest priority is given to maximizing the
number of users served by the HAPS--RIS setup. This is due to the fact that the HAPS--RIS is introduced as an assisting non-terrestrial component whose primary purpose is to enhance the existing UAV-based network.
Giving the first priority to $\tilde{U}^{\rm HAPS}$ ensures that the optimization fully exploits the HAPS--RIS capability before relying on UAV resources. 
The second priority is given to minimizing the number of required UAVs.
Compared to the total average UAV path loss, the limitation on the number of UAVs
constitutes a significantly more dominant practical challenge. UAVs are costly to deploy and operationally demanding, and hence, reducing $N^{\rm UAV}$ directly improves system feasibility. For this reason, among all solutions that achieve the same HAPS--RIS coverage level, the framework prioritizes those that require fewer UAVs, even if the corresponding UAV path loss is slightly higher. Finally, the third objective minimizes the average UAV path loss upper bound. Overall, this lexicographic ordering reflects the practical deployment logic of
the system: first maximize the benefit provided by the HAPS--RIS, then minimize the dependence on the limited UAV numbers, and subsequently refine the UAV communication performance. The output of the Pareto block for each value of $\varkappa$ is passed to the convergence check, and if the criterion is not met, the bandwidth portioning factor is updated accordingly. This iterative process continues until the convergence condition is satisfied. Thus, the solution of OP~4 emerges from an iterative two-layer procedure: an outer loop that adjusts the coupling variable $\varkappa$, and an inner
Pareto-based lexicographic optimizer that resolves the prioritized objectives
for each $\varkappa$.

\makeatletter
\renewcommand{\thealgorithm}{}
\makeatother
\begin{algorithm}[h]
\floatname{algorithm}{\textcolor{black}{The Proposed Dynamic Pareto Optimization Method:}}
\caption{ Solving \textbf{OP 4}}
\label{alg:OP4}

\textbf{s1:} Initialize $t=0$, $\delta=50$, $\delta'$=1, $N^{\rm{UAV},(0)}$, $R^{*}=R_{0}$, and define the set $\mathcal{I}$=\{$i|\sqrt{x^{2}_{i}+y^{2}_{i}} \leqslant R_{0}$\}. Initialize index $q=1$, maximum iteration $Q_{\max}$,
define the discrete bandwidth portioning set 
$\mathcal{\varkappa}=\{\,1,\,1.5,\,2,\,2.5,\,\ldots\,\}$ where $\varkappa^{(q)} = \mathcal\varkappa[q]$.

\textbf{s2:} \textbf{while} $q < Q_{\max}$ \textbf{do}

\textbf{s3:} \textit{Maximizing $\tilde{U}^{\rm{HAPS}}$ }
  
       {\textbf{s3-I:}}~ \textbf{for} $t=1$ to $T$ {\textbf{do}}
		\\ \hspace{0.5cm} {\textbf{s3-II:}}~ Define the set ${\mathcal{B}^{(t)}}$=\{$i'|\sqrt{x^{2}_{i'}+y^{2}_{i'}} \leqslant R^{(t)}$\} where $R^{(t)}=R_{0}-t\delta$. \textcolor{black}{$T$ is the upper limit of the iteration count}. Define the set $\mathcal{C}^{(t)}= \mathcal{I}-\mathcal{B}^{\left( t\right)}$ where the users in this area are supported by HAPS-RIS, and then calculate the number of the RIS elements per each cluster as $\left\lfloor\frac{M}{\left| \mathcal{C}^{(t)}\right|}\right\rfloor $.   
 
  {\textbf{s3-III:}}~ Map $\left| \mathcal{C}^{(t)}\right|$ clusters of RIS elements to $\left| \mathcal{C}^{(t)}\right|$  users based on a random injective function. Allocate $L^{(t)}_{\rm{Cluster}}=\lfloor\frac{L^{\rm{CS}}}{\left| \mathcal{C}^{(t)}\right|}\rfloor$ subcarriers to each RIS cluster. Then, uniformly allocate $L^{(t)}_{\rm{Cluster}}$ subcarriers to $\left\lfloor\frac{M}{\left| \mathcal{C}^{(t)}\right|}\right\rfloor$ RIS elements for each RIS cluster and calculate $\boldsymbol{\rho}(R^{(t)})$.
   
   {\textbf{s3-IV:}}~\textbf{While} ~~$r^{\rm{CS} }_{i}\left( \boldsymbol{\rho} \left( R^{(t)}\right)  \right)  \geq r^{\rm{user} }_{0}$, $\forall i\in\mathcal{C}$ and the HAPS-RIS zone constraints are met and $t < T$,  set $R^{*}=R^{(t)}$. Set $t=t+1$.
        \\~~~{\textbf{s3-V:}}~{\textbf{end for}}
		
\textbf{s4:} \textit{Minimizing $N^{\rm{UAV}}$ while minimizing $\varLambda ^{\rm{UAV,upp}}$}  
\\{\textbf{s4-I:}} Define the set ${\mathcal{B^{*}}}$=\{$i'|\sqrt{x^{2}_{i'}+y^{2}_{i'}} \leqslant R^{*}$\} where $R^{*}$ is obtained in S3.
  
   {\textbf{s4-II:}}~ \textbf{for} $t'=1$ to $T'$ {\textbf{do}}
		
{\textbf{s4-III:}} Obtain $\boldsymbol{x}\left( N^{\text{UAV},(t')} \right)$, $\boldsymbol{y}\left( N^{\text{UAV},(t')} \right)$, and the user association part of $\boldsymbol{\tilde{\rho}}\left( N^{\text{UAV},(t')} \right)$ based on the $k$-means clustering method for the users as members of the set $\mathcal{B}^{*}$, where $N^{\rm{UAV},(t')} = N^{\rm{UAV},(0)} - (t'-1)\delta'$. Apply orthogonality check for subcarriers based on the constraint $\sum\limits^{I}_{i} \tilde\rho^{l'}_{ij} \leq 1\,\forall l',j$. Then calculate $\boldsymbol{\tilde\rho} \left( N^{\text{UAV},(t') }\right)$.
   
   {\textbf{s4-IV:}}~\textbf{While} ~~ $t' < T'$ and
  $r^{\rm{UAV} }_{i}\left( \boldsymbol{\tilde\rho} \left( N^{\text{UAV},(t') }\right), \boldsymbol{x}\left( N^{\text{UAV},(t') }\right),\boldsymbol{y}\left( N^{\text{UAV},(t') }\right) \right) \geq r^{\rm{user} }_{0},\  \forall i \in\mathcal{B}$ and the UAV zone related constraints are met,  set $N^{*}=N^{{\rm{UAV}},(t')}$. Set $t'=t'+1$.
        \\{\textbf{s4-V:}}~{\textbf{end for}}

\textbf{s5:} \textbf{if} $|N^{\rm{UAV} (q+1)}-N^{\rm{UAV} (q)}|<\varepsilon$ and $|\tilde{U}^{\rm{HAPS} (q+1)} -\tilde{U}^{\rm{HAPS} (q)} |<\varepsilon $
\textbf{then break}; \textbf{else} set $q\leftarrow q+1$.

\textbf{s6:} The solution of \textbf{OP 4} is obtained.

\end{algorithm}

\vspace{1mm}

\textit{\textbf{Corollary 1:}}
Through the principled sequence of analytical transformations in \emph{Theorem~1} and \emph{Lemma~1} and the proposed mapping technique in \emph{Proposition~1}, the intractable optimization problem \textbf{OP~1} is relaxed and mapped into a tractable optimization problem, \textbf{OP~4}. The proposed geometric model of \textbf{OP~4} is depicted in Fig. \ref{sys2}. In \emph{Theorem~1}, $k$-means clustering is exploited to minimize the upper bound of \eqref{third obj}, and hence \textbf{OP~1} can be relaxed into \textbf{OP~2}. In \emph{Lemma~1}, a practical closed-form RIS phase-shift design is exploited to relax \textbf{OP~2} into \textbf{OP~3}. Finally, \textbf{OP~3} is mapped into \textbf{OP~4} via \emph{RIS clustering}, \emph{BW portioning}, and \emph{zone association}. The solution of \textbf{OP~4} can be obtained via the proposed dynamic Pareto optimization technique where the building blocks and the interfaces are illustrated in Fig. \ref{sys3}.
\vspace{-2mm}
\section{Complexity and Convergence Analysis}
In this section, we prove the low computational complexity and guaranteed convergence of the proposed dynamic Pareto optimization method, which are critical for ensuring scalable and reliable deployment.
\vspace{-5mm}
\subsection{Computational Complexity}
In the proposed dynamic Pareto optimization technique, for each candidate value of the bandwidth portioning factor~$\varkappa$, up to $Q_{\max}$ candidates, the method 
executes two main blocks. 
Each operation represents a constant-cost algorithmic task whose execution complexity is independent of the system size and is therefore treated as $\mathcal{O}(1)$. For a fixed $\varkappa$, the HAPS--RIS block performs at most $T$ updates of the UAV-zone radius. In the worst case, each update involves partitioning all users into HAPS--RIS and UAV zones, which requires $\mathcal{O}(I)$ operations; forming RIS clusters and mapping them to HAPS--RIS users, incurring $\mathcal{O}(I)$ operations; allocating CS subcarriers and RIS elements across clusters, with  $\mathcal{O}(L^{\rm CS}(\varkappa)+M)$ operations; and verifying rate and feasibility constraints for all HAPS--RIS users with $\mathcal{O}(I)$ operations. Consequently, the worst-case computational complexity per update scales as $\mathcal{O}(I+L^{\rm CS}(\varkappa)+M)$, yielding a computational complexity for the HAPS--RIS block as
\begin{equation}
    \mathcal{O}\!\left(T(I+L^{\rm CS}(\varkappa)+M)\right).
\end{equation}
The UAV block performs at most $T'$ UAV-reduction steps. In the worst case, each step clusters all UAV-zone users using the $k$-means algorithm to update UAV locations and user associations. Since the clustering processes all $I$ users across the initial $N_{\rm UAV}^{(0)}$ UAVs and requires at most $T_k$ internal iterations, its complexity scales as $\mathcal{O}(I N_{\rm UAV}^{(0)} T_k)$. In addition, each step verifies rate and feasibility constraints for UAV-served users over $L^{\rm UAV}(\varkappa)$ subcarriers, contributing an extra $\mathcal{O}(I N_{\rm UAV}^{(0)} L^{\rm UAV}(\varkappa))$ operations. Combining these two components and accounting for $T'$ reduction steps yields a computational complexity for the UAV block as
\begin{equation}
    \mathcal{O}\!\left(T' I N_{\rm UAV}^{(0)} (T_k + L^{\rm UAV}(\varkappa))\right).
\end{equation}
To obtain the worst-case overall computational complexity, we apply the following bounds:  
(i) the subcarrier counts satisfy $L^{\rm CS}(\varkappa)\le L^{\rm tot}$ and 
$L^{\rm UAV}(\varkappa)\le L^{\rm tot}$ for all tested values of ~$\varkappa$;  
(ii) the number of users in each zone is bounded by $|B|\le I$ and $|C|\le I$;  
(iii) the number of UAVs during the reduction steps satisfies 
$N_{\rm UAV}^{(t')}\le N_{\rm UAV}^{(0)}$; and  
(iv) the algorithmic iteration limits $T$, $T'$, $T_k$, and $Q_{\max}$ are finite. Accordingly, the overall computational complexity can be obtained as
\begin{equation}
    \mathcal{O}\!\Big(
Q_{\max}\big[
T(I + L^{\rm tot} + M) + T' I N_{\rm UAV}^{(0)}(T_k + L^{\rm tot})
\big]
\Big).
\end{equation}
All loops are bounded and every operation grows only polynomially with the system dimensions. Therefore, the proposed solution is a low-complexity and scalable approach.
\vspace{-3mm}
\subsection{Convergence Analysis}
For a fixed value of $\varkappa$, the proposed method first scans the UAV-zone radius as
$R(t)=R_0-t\delta$, for $t=1,\dots,T$, over a finite discrete grid. This scan is finite and sequence $\{R(t)\}$ is decreasing monotonically which enlarges the HAPS--RIS user set. Consequently, the HAPS coverage
metric $\tilde U^{\rm{HAPS}}$ is monotonically non-decreasing along the scan.
The algorithm stores the smallest radius that satisfies the HAPS--RIS feasibility
constraints, which maximizes $\tilde U^{\rm{HAPS}}$ over the tested radius
grid. Given this radius, the proposed technique then scans the number of UAVs as
\begin{equation}
    N_{\rm{UAV}}^{(t')}=N_{\rm{UAV}}^{(0)}-(t'-1)\delta', ~ \forall t' \in \{1,\dots,T'\}.
\end{equation}
This scan is also finite and monotonic. The smallest UAV count
that satisfies the UAV-zone constraints is retained, yielding the minimum feasible $N_{\rm{UAV}}$. For each tested
UAV count, the $k$-means-based UAV placement monotonically decreases a bounded
objective and therefore terminates in finite time. Finally, the outer loop evaluates $\varkappa$ over the finite set 
and is capped by $Q_{\max}$. Hence, the algorithm explores only a finite number of candidate solutions and necessarily terminates. Since updates
are accepted only when the corresponding feasibility conditions are satisfied,
the final output is feasible for \textbf{OP 4}. In summary, the proposed method implements a dynamic lexicographic Pareto search over finite grids, which guarantees convergence.

\vspace{-3mm}
 \section{Case Study and Numerical Analysis}

The numerical analysis are organized around three case studies. First, we consider a \emph{UAV-free case study}, which arises as a \emph{degenerate case} of \textbf{OP~4} by setting $N_{\rm UAV}=0$ and $\varkappa=1$. Under this configuration, the HAPS--RIS operates as a standalone tier, utilizing 50~MHz bandwidth without any UAV assistance. Second, we consider a \emph{UAV in action but separate spectrum case study} that arises as a disjoint degenerate case of \textbf{OP~4} by setting $\varkappa=1$. In this case, HAPS--RIS and UAVs operate in parallel with disjoint and dedicated bandwidths of 50~MHz each, enabling the evaluation of their inherent performance trade-offs. Finally, we investigate the \emph{UAV in action with shared spectrum case study} in which the full formulation of \textbf{OP~4} is activated. In this case, the entire 100~MHz bandwidth is treated as a shared resource and adaptively partitioned between the HAPS--RIS and UAV systems. For the proposed unified joint case study, to benchmark the framework and to highlight the impact of bandwidth coupling, we evaluate the system performance under four representative bandwidth allocation regimes. All schemes are characterized through the bandwidth portioning factor $\varkappa$, which enables a unified interpretation of the baseline strategies as special operating points of the proposed model. When $\varkappa_0 = \varkappa\rightarrow 0$, almost the entire available bandwidth is allocated to the UAVs, i.e., $L_{\rm{UAV}} \gg L_{\rm{CS}}$. In this regime, the contribution of the HAPS-RIS becomes negligible, and the system effectively operates as a \emph{UAV-only} network. When $\varkappa_1=\varkappa \rightarrow \infty$, nearly all bandwidth is assigned to the HAPS-RIS system, i.e., $L_{\rm{CS}} \gg L_{\rm{UAV}}$. This configuration corresponds to the \emph{HAPS-RIS-only} baseline. Between these two limiting cases, the operating point $\varkappa = 1$ represents an \emph{equal bandwidth allocation} baseline, where the total bandwidth is evenly divided between the HAPS-RIS and UAV infrastructures. Although this strategy enables simultaneous utilization of both technologies, it cannot fully exploit the complementary capabilities of the two aerial systems. Finally, $\varkappa = \varkappa^{\rm{opt}}$ denotes the operating point obtained via the proposed \emph{dynamic Pareto optimization framework}. This optimized regime captures the best trade-off between the HAPS-RIS and UAV infrastructures and serves as the primary benchmark for evaluating the performance gains of the proposed joint design over all baseline schemes.

In the simulations, the parameters of the HAPS--RIS system are selected based on
established channel models and system assumptions reported in
\cite{alfattani2021aerial, alfattani2021link}, while the UAV-related parameters
follow the settings adopted in \cite{al2014optimal, azizi2019profit}. This choice ensures a realistic and consistent evaluation of the proposed framework. We consider \(I=20\) ground users uniformly distributed, with a minimum separation distance of $D_{0}=$ 100 m among them, within a circular area of
radius \(R_0=500\)~m centered at \((x_0,y_0,z_0)=(0,0,0)\). The CS
is located at \((x^{\rm{CS}},y^{\rm{CS}},z^{\rm{CS}})=(-10000,0,1000)\),
and the HAPS platform is positioned at
\((x^{\rm{HAPS}},y^{\rm{HAPS}},z^{\rm{HAPS}})=(-5000,100,20000)\), in Cartesian coordinate system. As elaborated in Section \ref{section 2}, the HAPS--RIS channel is modeled as a cascaded CS--RIS--user link dominated by LoS propagation, while the UAV channel follows an air-to-ground propagation model where the channel gain depends on the average path loss including LoS and NLoS probabilities. The CS employs a directional antenna with gain \(G^{\rm{CS}}=43.2\)~dB, while
the antennas at the user terminals and UAVs are assumed to be omnidirectional
with gains \(G_i=0\)~dB and \(G_j=0\)~dB, respectively. Unless otherwise stated, the CS transmit power is fixed at \(40\)~dBm, while each
UAV transmits with a power of \(20\)~dBm. The path loss exponent is \(\alpha=2\), and the excessive path loss coefficients for LoS and NLoS links are \(\eta_1=1\) and \(\eta_2=31\), respectively. The environment-dependent
constants are chosen as \(\psi=5\) and \(\beta=0.5\). In the benchmarking analysis, we interpret the baseline strategies as specific operating points within the proposed model. The system operates at a carrier frequency of \(f_{\rm{c}}=2\)~GHz and utilizes \(L=64\) orthogonal subcarriers. Each RIS element is assumed to have unit reflection efficiency, i.e.,
\(\mu_m=1\) for all \(m\). The noise power spectral density is set to
\(N_0=-174\)~dBm/Hz.
\vspace{-4mm}
\subsection{UAV-free Case Study: A Degenerate Case of \textbf{OP 4}}
\begin{figure}[t]
\centering
\begin{minipage}[t]{0.5\linewidth}
    \centering
    \includegraphics[width=\linewidth]{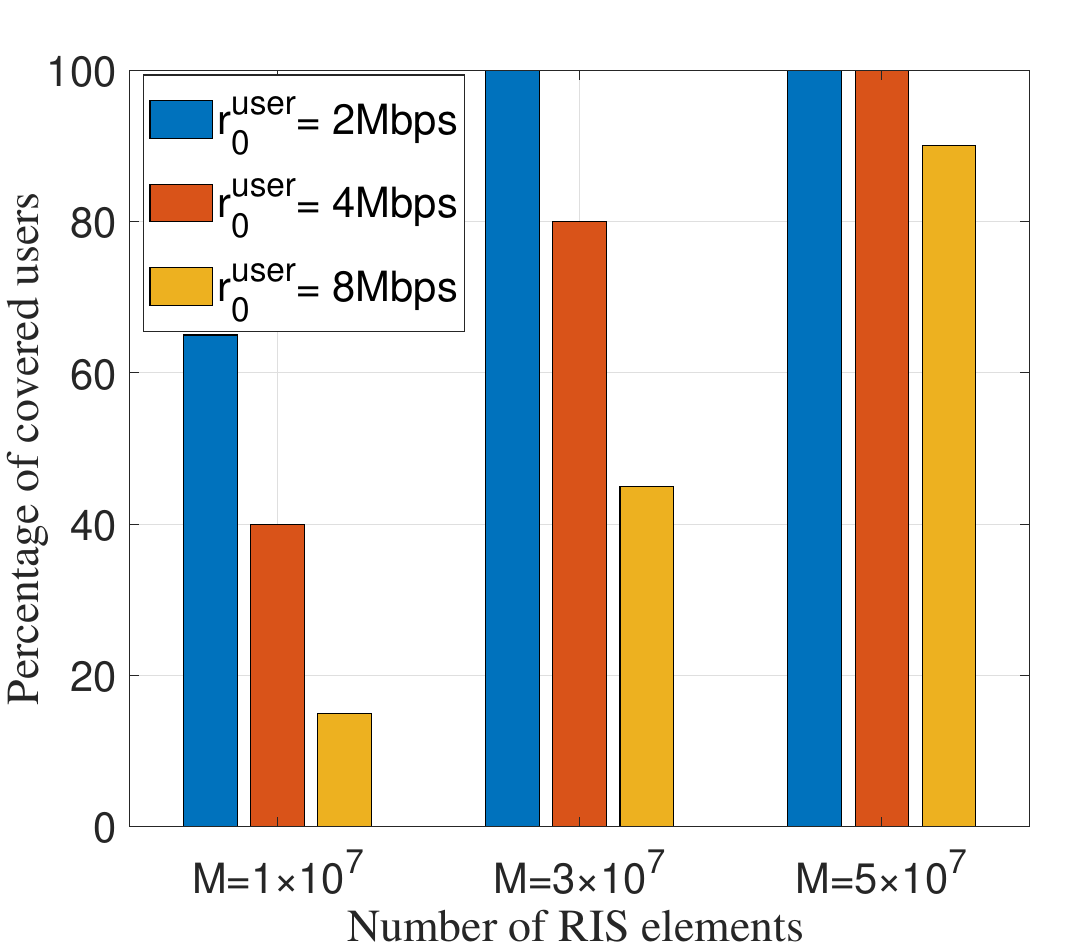}
    \caption*{(a) High rate regime}
\end{minipage}\hfill
\begin{minipage}[t]{0.5\linewidth}
    \centering
    \includegraphics[width=\linewidth]{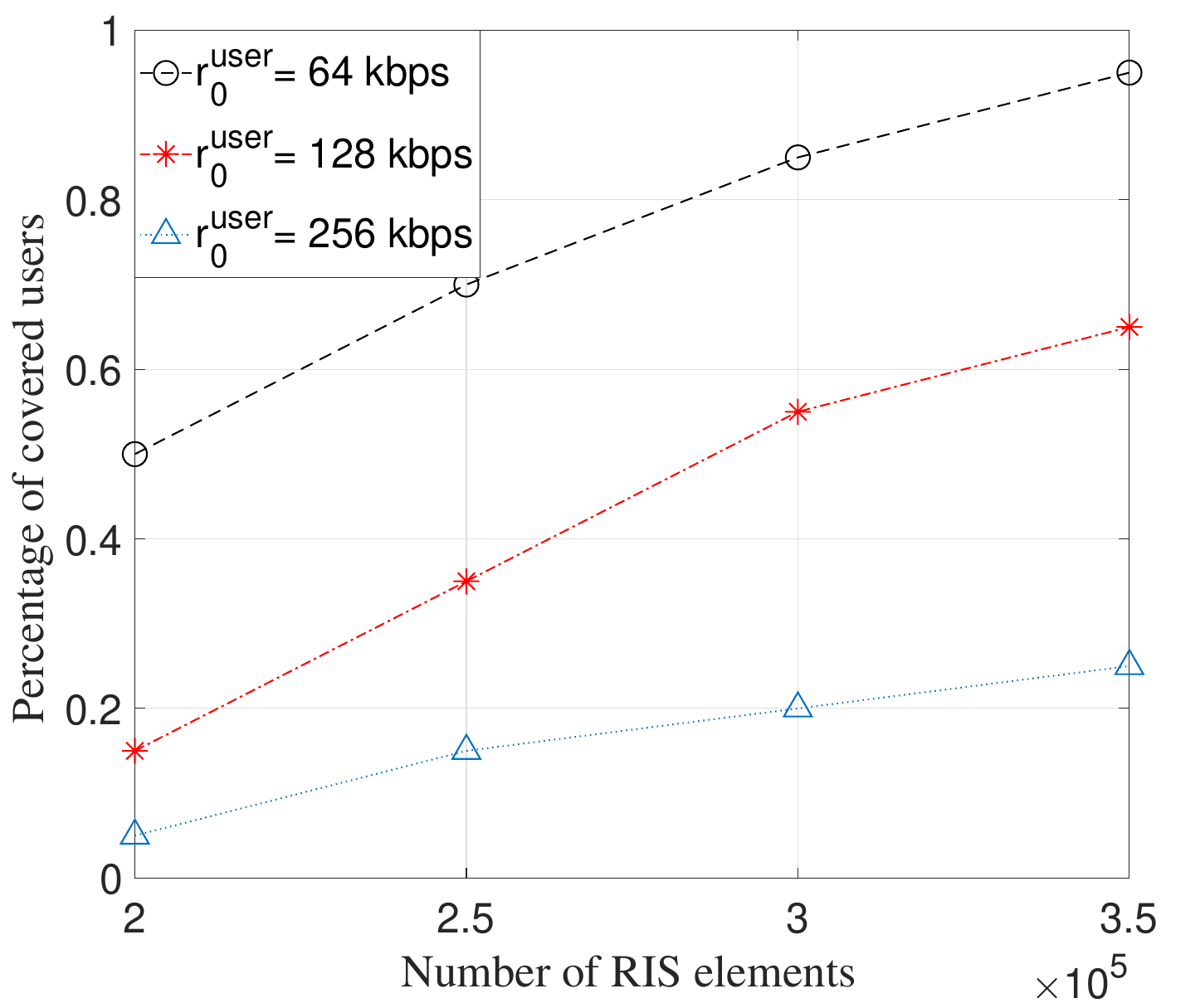}
    \caption*{(b) Low rate regime}
\end{minipage}
\caption{Percentage of covered users by HAPS-RIS versus number of RIS elements for the UAV-free case with the CS transmit power of 40~dBm.}
\label{percentUserHAPSonly}
\end{figure}

\begin{figure}[t]
    \vspace{-13mm}

\centering
\begin{minipage}[t]{0.5\linewidth}
    \centering
    \includegraphics[width=\linewidth]{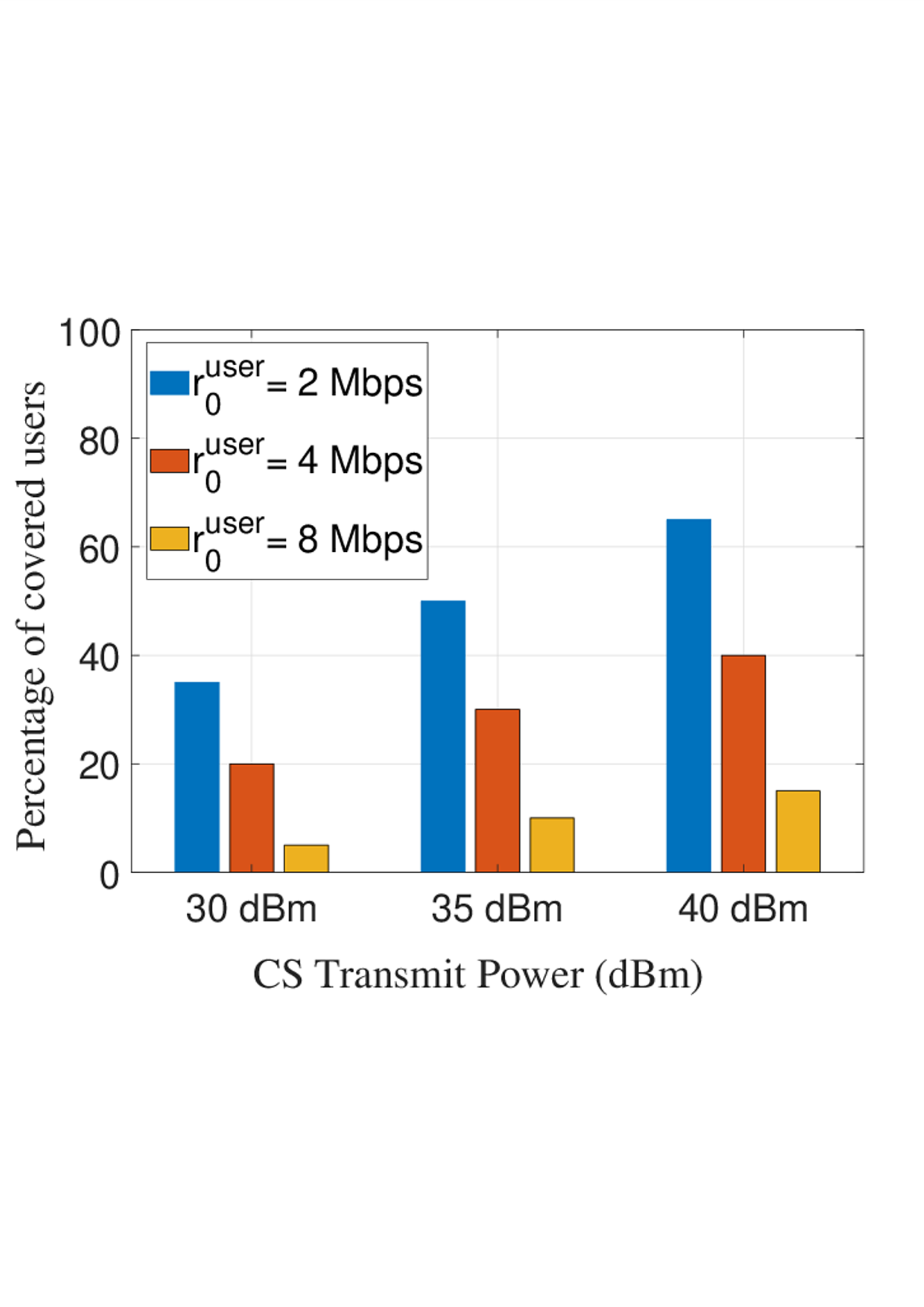}
    \caption*{(a) High rate regime with $10^7$ RIS elements in use.}
\end{minipage}\hfill
\begin{minipage}[t]{0.48\linewidth}
    \centering
    \includegraphics[width=\linewidth]{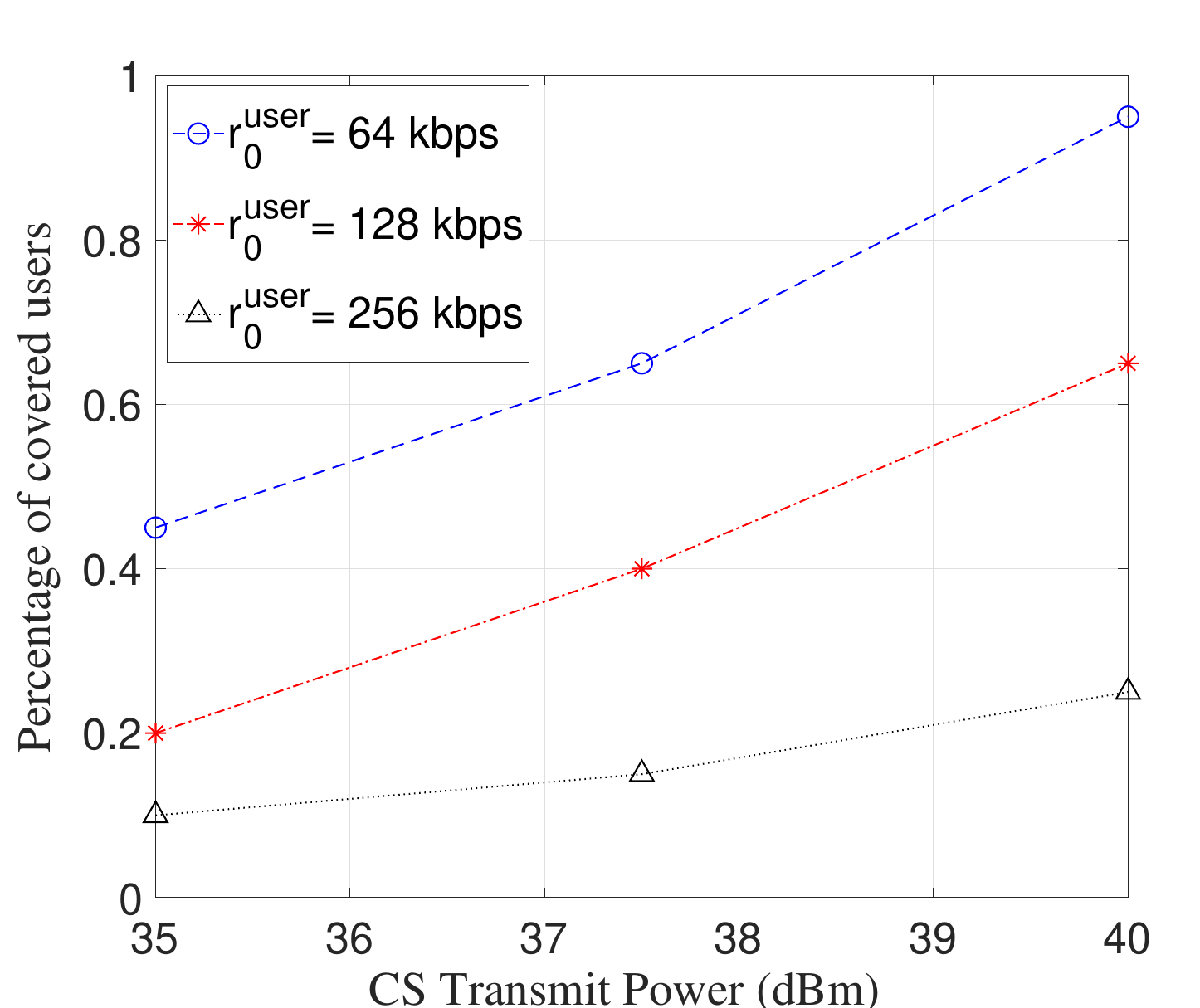}
    \caption*{(b) Low rate regime with $3.5\times10^5$ RIS elements in use.}
\end{minipage}
\caption{Percentage of covered users by HAPS-RIS versus CS transmit power for the UAV-free case.}
\label{percentageUserHAPSonly-CS}
\end{figure}

We consider a UAV-free case study, obtained as a degenerate case of \textbf{OP~4} by setting $N_{\rm UAV}=0$ and $\varkappa=1$, in which the HAPS--RIS operates as a standalone tier using a dedicated 50~MHz bandwidth without UAV assistance.
Fig.~\ref{percentUserHAPSonly} shows the percentage of users covered by the HAPS-RIS versus the number of RIS elements under different rate requirements, considering a HAPS-RIS-only deployment. In the low-rate regime, near-full coverage is achieved with a relatively small number of RIS elements. In contrast, in the high-rate regime, coverage improves more gradually and requires a substantially larger RIS surface to satisfy the increased SNR requirements. 
Fig.~\ref{percentageUserHAPSonly-CS} illustrates the percentage of users covered by the HAPS-RIS as a function of the CS transmit power under different rate requirements. In the low-rate regime, high coverage is achieved even at moderate transmit power levels, indicating that the HAPS-RIS can efficiently satisfy modest rate demands with limited CS power. In contrast, in the high-rate regime, coverage becomes strongly power-limited, and a substantial increase in CS transmit power is required.

\subsection{UAV in Action but Separate Spectrum Case Study: A Disjoint Degenerate Case of \textbf{OP 4}}

\begin{figure}[h]
\centering
\includegraphics[scale=0.15]{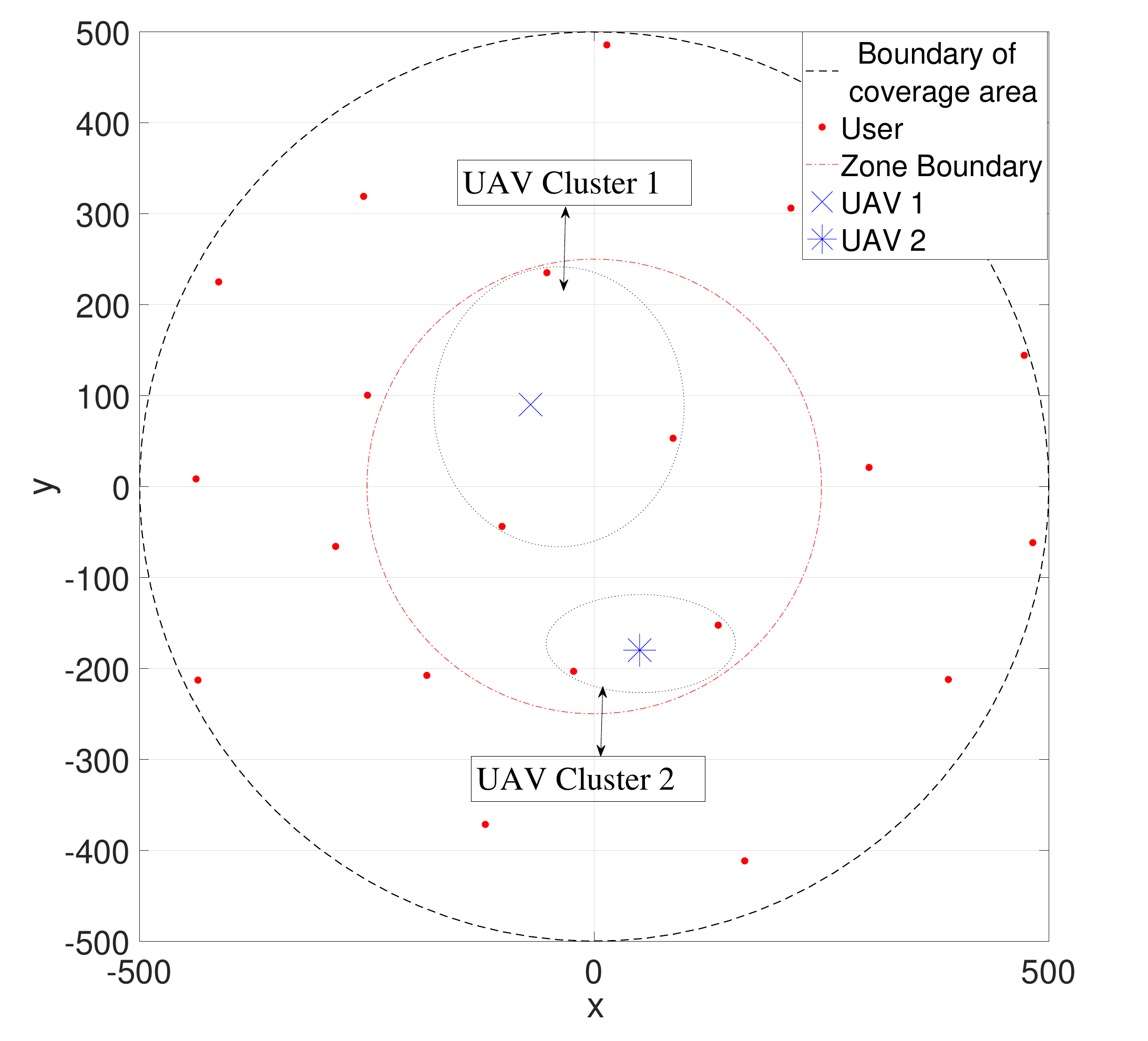}
      \caption{An example of user association illustrating the coexistence of HAPS-RIS and UAVs via zone-based association and UAV clustering.}
    \label{Snapshot}
\end{figure}

\begin{figure}
\centering
  \includegraphics[scale=0.15]{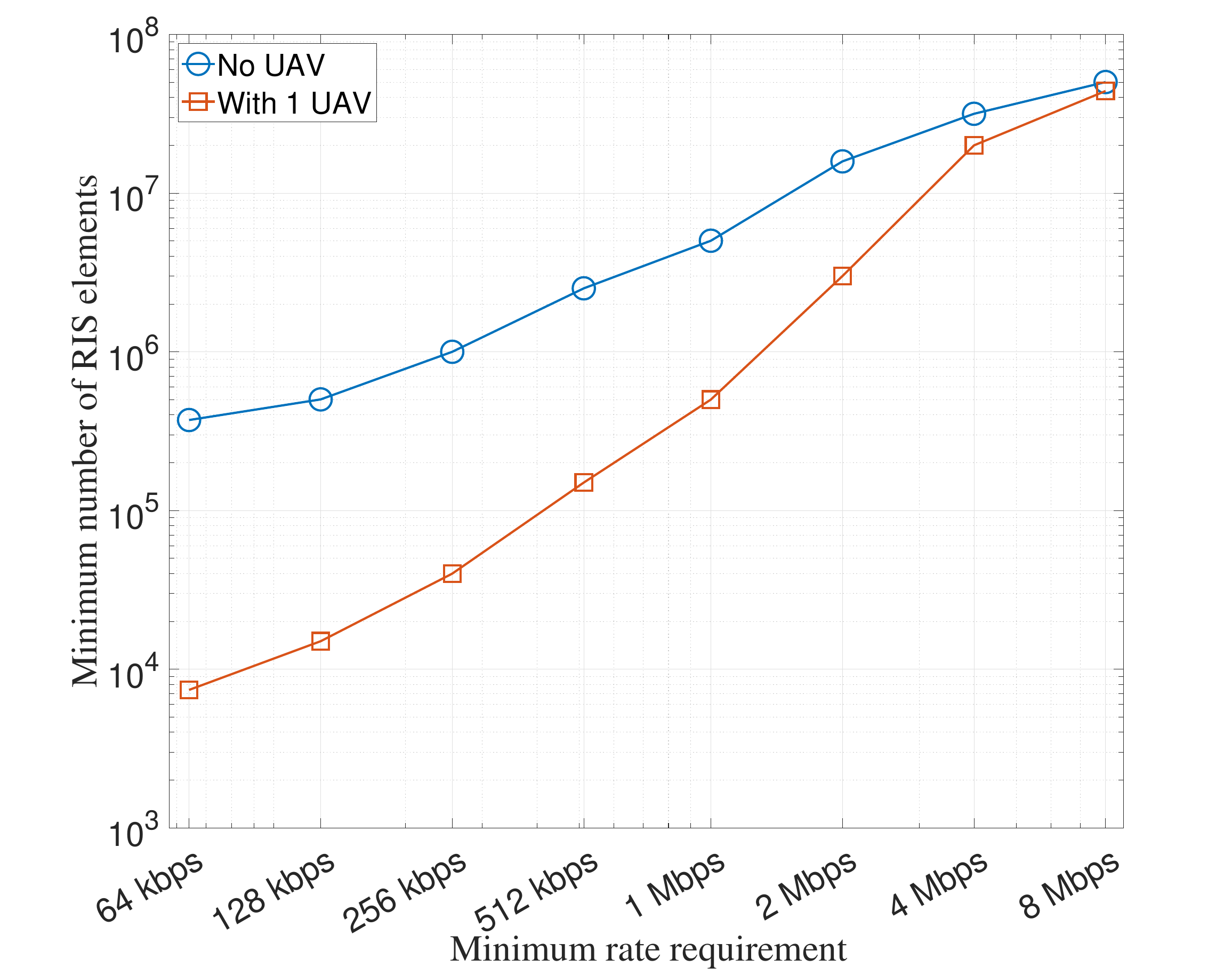}
  \caption{Minimum Required number of total RIS elements  versus the wide range of rate requirements to provide full coverage.}
    \label{logMr}
\end{figure}

In this case, we consider a UAV-assisted case study with separate spectrum allocation, which arises as a disjoint degenerate case of \textbf{OP~4} by fixing $\varkappa=1$. In this configuration, the HAPS--RIS and UAV tiers operate in parallel over dedicated 50~MHz bandwidths, enabling an assessment of their inherent performance trade-offs. Fig.~\ref{Snapshot} illustrates an example of this scenario through the zone-based association and UAV clustering mechanism. The coverage area is partitioned by a zone boundary into two distinct regions: an outer HAPS-RIS zone, where users are served by the HAPS-RIS system, and an inner UAV zone, where users are supported by UAVs. This spatial separation enables the two infrastructures to operate simultaneously while serving different subsets of users without interference. Within the UAV zone, users are further grouped into clusters based on the $k$-means clustering approach, with each cluster being served by a dedicated UAV positioned at the corresponding cluster centroid. 
Fig.~\ref{logMr} depicts the minimum number of RIS elements required to achieve full coverage over a wide range of user rate requirements, spanning from low-rate to high-rate regimes. In the low-rate regime, exploiting a single UAV leads to a significant reduction in the required RIS size. In contrast, as the rate requirement increases, the performance gap between the HAPS-RIS-only and UAV-assisted cases progressively narrows. Notably, in the high-rate regime, particularly for rate requirements exceeding $4$~Mbps, the assistance provided by a single UAV becomes marginal, and deploying multiple UAVs is necessary to substantially reduce the required number of RIS elements and provide the full coverage.

\begin{figure}[t]
\centering
\begin{minipage}[t]{0.5\linewidth}
    \centering
    \includegraphics[width=\linewidth]{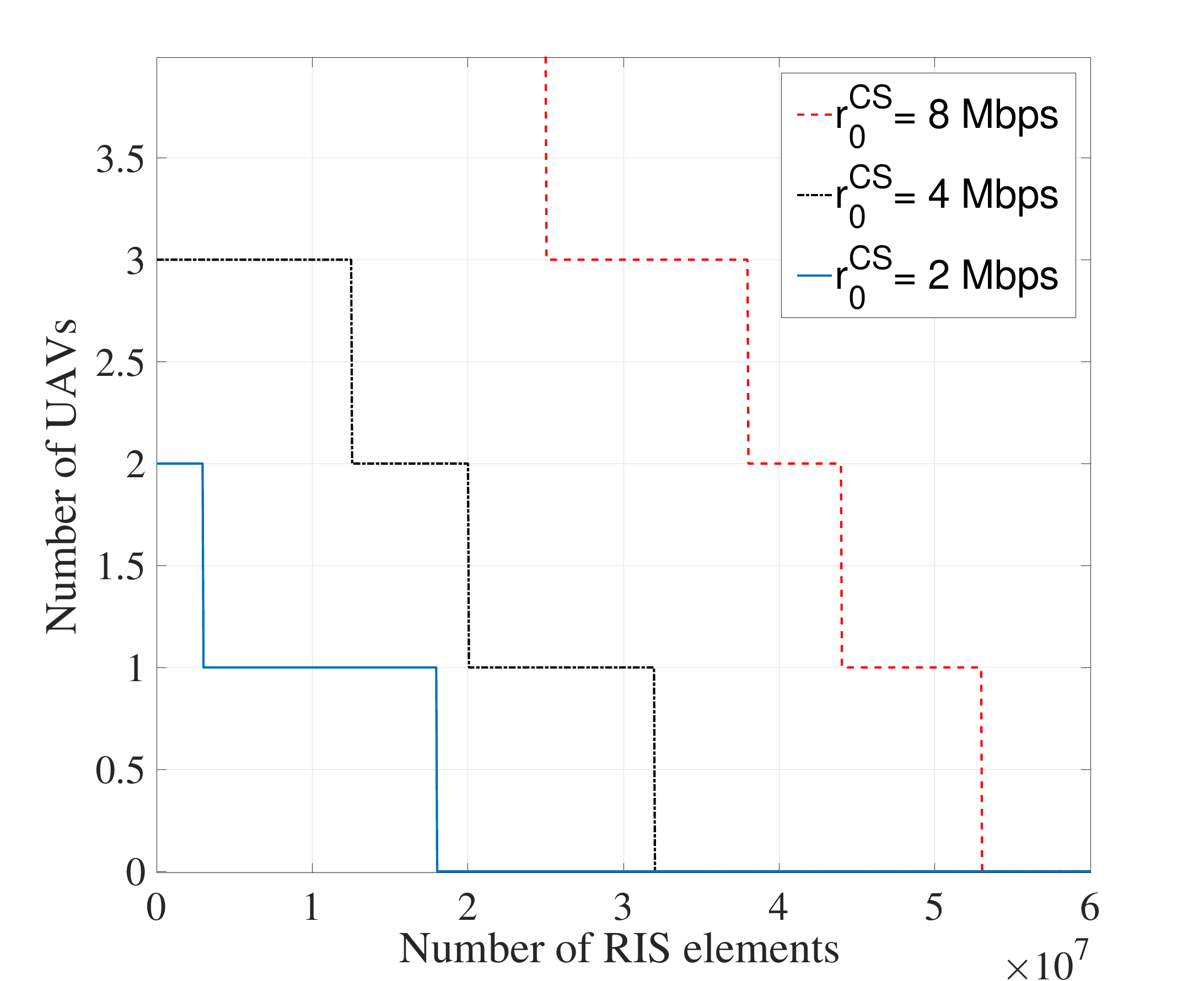}
    \caption*{(a) High rate regime}
\end{minipage}\hfill
\begin{minipage}[t]{0.5\linewidth}
    \centering
    \includegraphics[width=\linewidth]{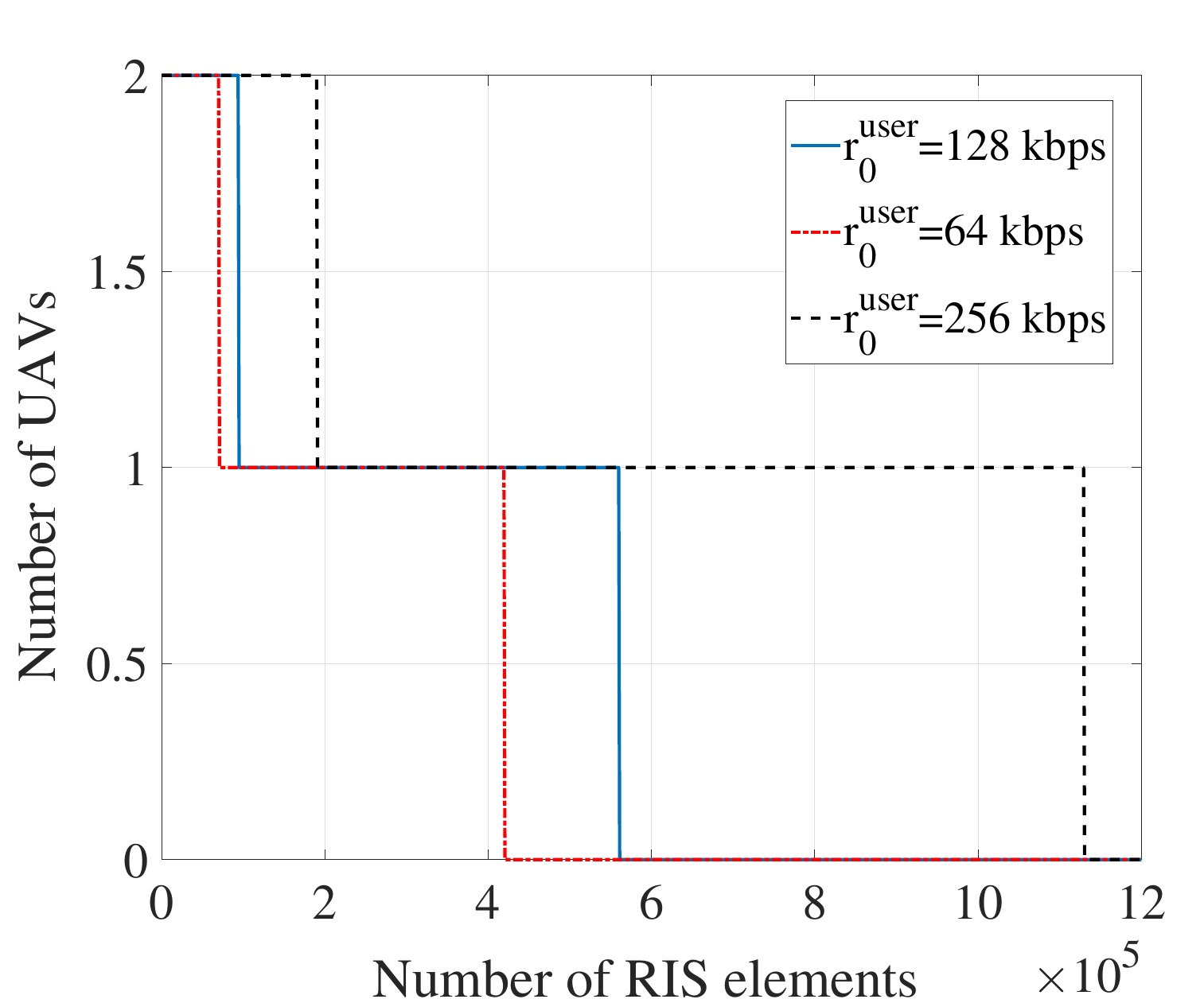}
    \caption*{(b) Low rate regime}
\end{minipage}
\caption{Trade-off between the minimum number of required UAVs and the number of RIS elements to provide full coverage.}
\label{NversusM}
\end{figure}

 Fig.~\ref{NversusM} depicts the required number of UAVs as a function of the total number of RIS elements in both high- and low-rate regimes. In the high-rate regime, the number of UAVs decreases in a stepwise fashion as $M$ increases, indicating that larger RIS sizes can progressively replace UAVs; however, this replacement typically requires RIS sizes on a much larger scale than in the low-rate case. In the low-rate regime, full coverage can be achieved with few UAVs even for moderate RIS sizes, and UAVs are eliminated with comparatively smaller increases in $M$, i.e., the number of RIS elements. This demonstrates a rate-dependent trade-off between RIS scale and UAV density.

\vspace{-4mm}
\subsection{UAV in Action with Shared Spectrum: Unified Joint Case Study and Comparison with its Baselines}
In this case study, we investigate the shared-spectrum UAV-assisted case study, in which the full formulation of \textbf{OP~4} is activated and the entire 100~MHz bandwidth is jointly and adaptively partitioned between the HAPS--RIS and UAVs. The system performance is benchmarked across four representative bandwidth regimes, all characterized by the bandwidth portioning factor~$\varkappa$. The limiting cases $\varkappa\!\to\!0$ and $\varkappa\!\to\!\infty$ recover the \emph{UAV-only} and \emph{HAPS--RIS-only} baselines, respectively, while $\varkappa=1$ corresponds to equal spectrum sharing. The operating point $\varkappa=\varkappa^{\rm opt}$, obtained via the proposed dynamic Pareto optimization, captures the optimized bandwidth portioning factor. 

\begin{figure}[t]
    \vspace{-11mm}
\centering
\begin{minipage}[t]{0.5\linewidth}
    \centering
    \includegraphics[width=\linewidth]{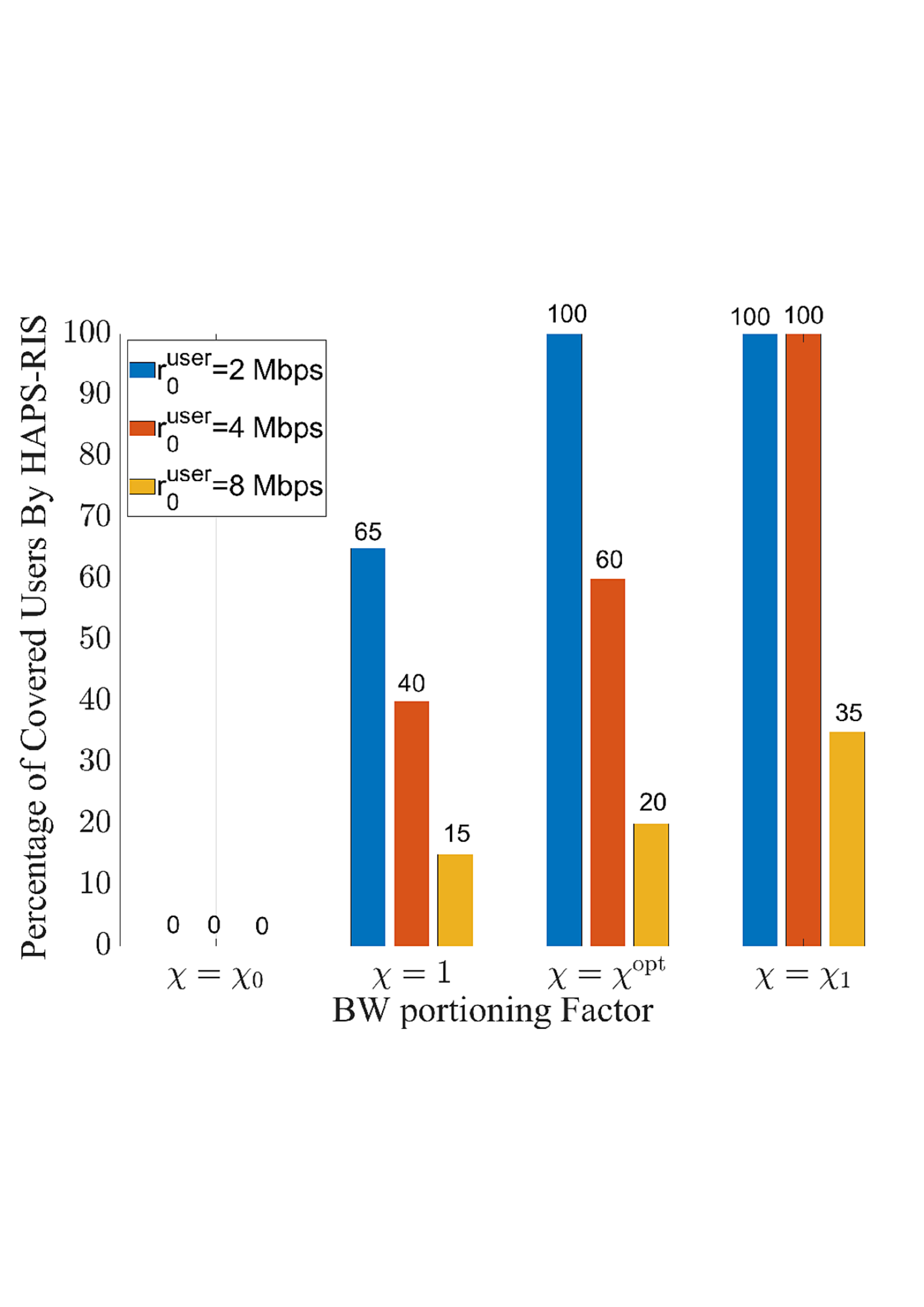}
    \caption*{(a) High rate regime where $10^7$ RIS elements are exploited.}
\end{minipage}\hfill
\begin{minipage}[t]{0.5\linewidth}
    \centering
    \includegraphics[width=\linewidth]{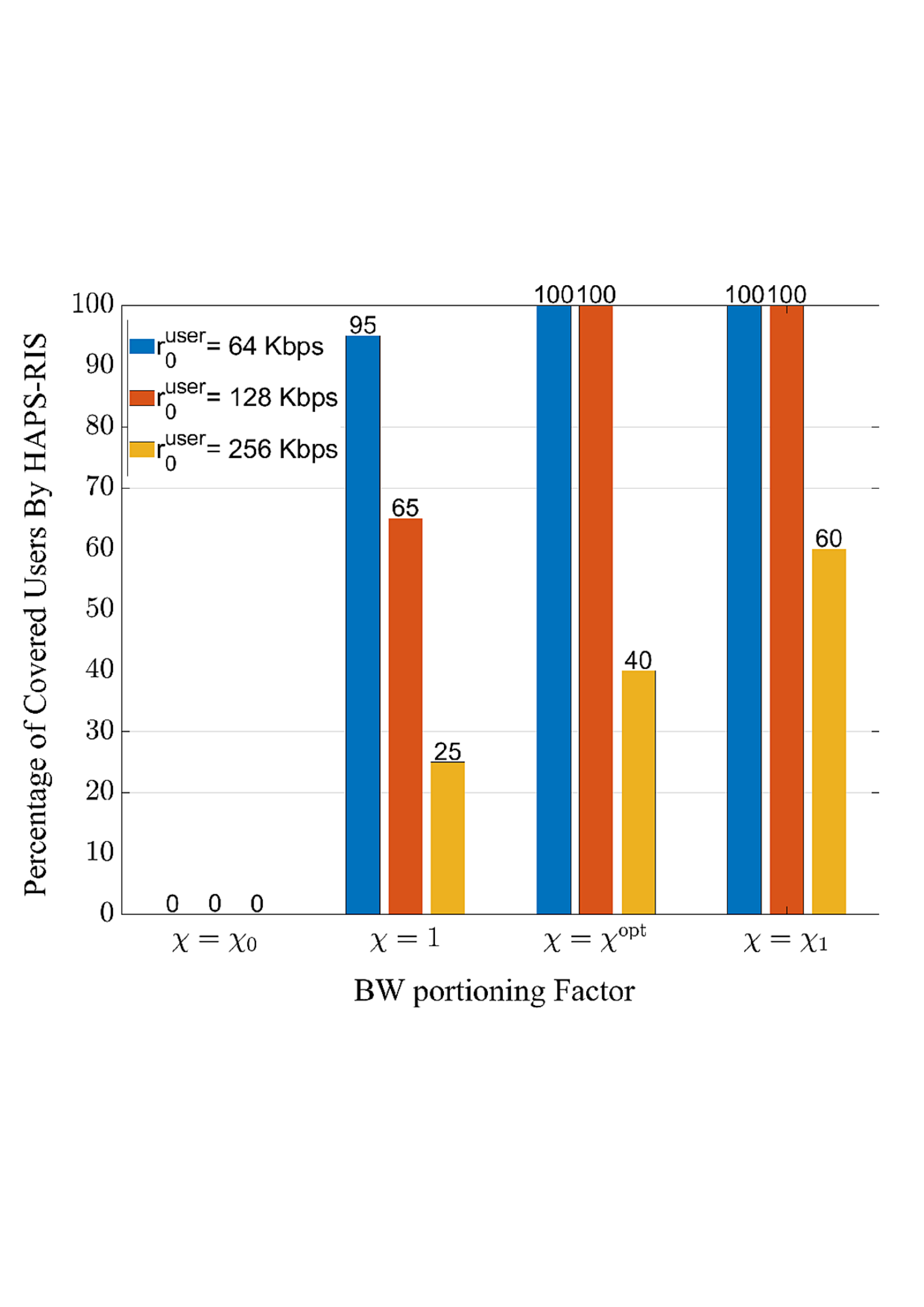}
    \caption*{(b) Low rate regime where $3.5\times10^5$ RIS elements are exploited.}
\end{minipage}

\caption{Percentage of covered users by HAPS-RIS versus BW portioning factor.}
\label{percentageUSerBaselines}
\end{figure}

Fig.~\ref{percentageUSerBaselines} compares the proposed unified joint solution of \textbf{OP 4}, $\varkappa=\varkappa^{\rm opt}$, with the other baselines which can be obtained by special cases of \textbf{OP 4} by setting $\varkappa\!\to\!0$, $\varkappa=1$, and $\varkappa\!\to\!\infty$. In the UAV-only case, where $\varkappa=\varkappa_0$, almost the entire bandwidth is allocated to the UAVs, leaving no resources for the HAPS--RIS and resulting in zero HAPS--RIS user coverage. The equal bandwidth allocation baseline, $\varkappa=1$, enables partial HAPS--RIS utilization; however, its performance is insufficient for high-rate demands. Allocating the whole bandwidth to the HAPS--RIS, $\varkappa = \varkappa_1$, improves HAPS--RIS coverage; however, it leads to outage in a case where users that cannot be supported by the HAPS--RIS. This is due to the high altitude of HAPS-RIS which can affect the path loss and rate. Finally, the proposed optimized allocation, where $\varkappa=\varkappa^{\rm{opt}}$, the bandwidth portioning factor is optimized by the proposed dynamic Pareto optimization technique. The optimized bandwidth allocation $\varkappa^{\rm{opt}}$ significantly improves the number of users covered by the HAPS--RIS compared to equal bandwidth allocation, while preserving the flexibility to allocate the remaining bandwidth to the UAVs. This regime enables the system to prioritize HAPS--RIS coverage without exhausting all resources while retaining the ability to serve any users left uncovered through UAV assistance and ultimately support full network coverage. A direct comparison between Fig.~\ref{percentUserHAPSonly} and Fig.~\ref{percentageUSerBaselines} is meaningful only under matched RIS configurations. Specifically, the high-rate results are comparable when $M=10^{7}$, while the low-rate results correspond to $M=3.5\times10^{5}$. Under these aligned conditions, Fig.~\ref{percentageUSerBaselines} exhibits a higher percentage of covered users than Fig.~\ref{percentUserHAPSonly}. In Fig.~\ref{percentUserHAPSonly}, the HAPS--RIS operates with a dedicated bandwidth of 50~MHz, whereas Fig.~\ref{percentageUSerBaselines} considers a shared bandwidth scenario in which, at $\varkappa=\varkappa_1$, the entire 100~MHz bandwidth is allocated to the HAPS--RIS setup. Consequently, the increased bandwidth availability significantly enhances HAPS-RIS user coverage. Hence, the higher coverage observed in Fig.~\ref{percentageUSerBaselines} is a direct outcome of bandwidth expansion.

\begin{figure}[t]
\centering
\begin{minipage}[t]{0.5\linewidth}
    \centering
    \includegraphics[width=\linewidth]{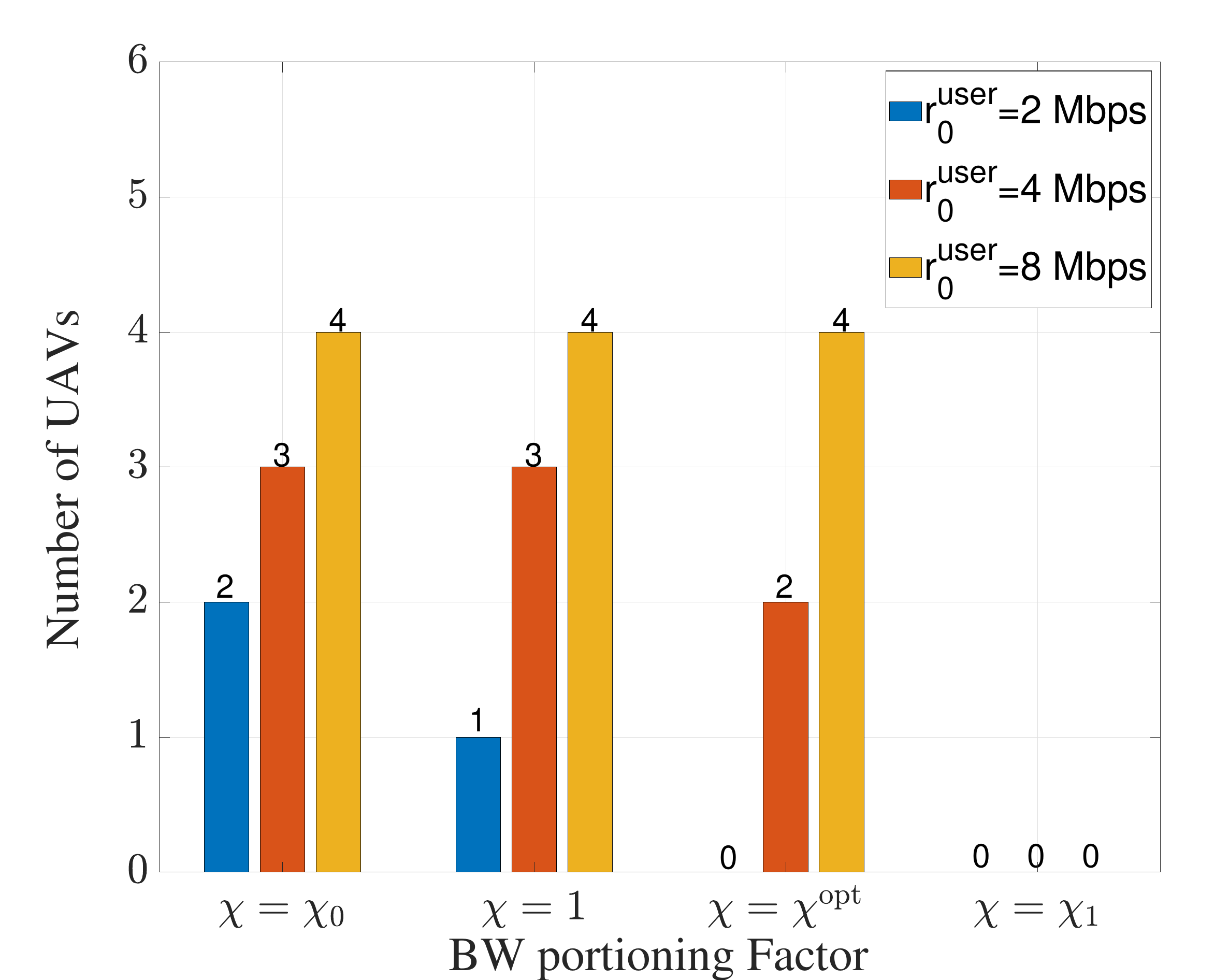}
    \caption*{(a) High rate regime where $10^7$ RIS elements are exploited.}
\end{minipage}\hfill
\begin{minipage}[t]{0.5\linewidth}
    \centering
    \includegraphics[width=\linewidth]{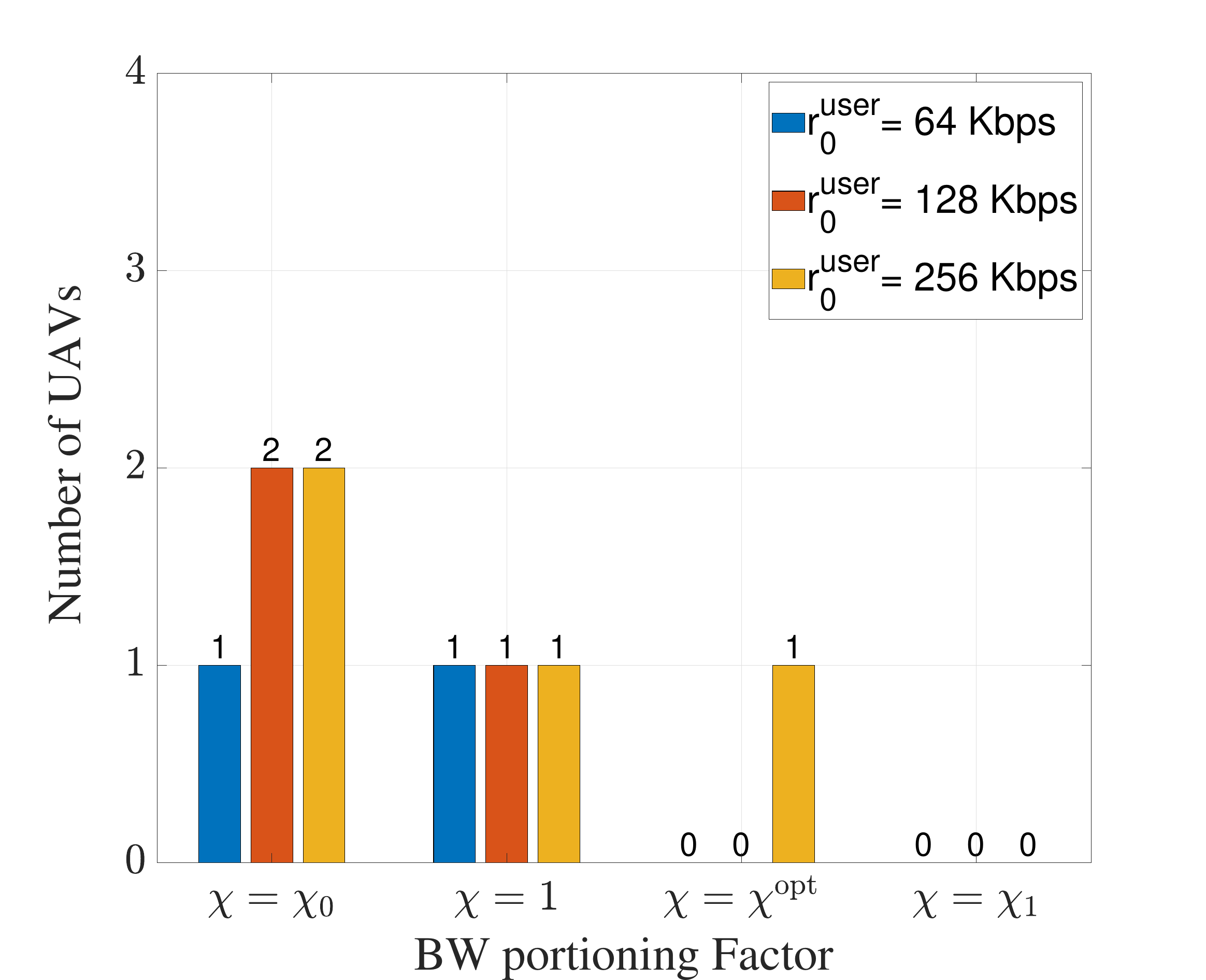}
    \caption*{(b) Low rate regime where $3.5\times10^5$ RIS elements are exploited.}
\end{minipage}
\caption{Number of UAVs versus BW portioning factor.}
\label{UAVNumberBaselines}
\end{figure}

Fig.~\ref{UAVNumberBaselines} illustrates the required number of UAVs under different bandwidth allocation regimes, revealing distinct behaviors in the low-rate and high-rate regimes. In both the high-rate and low-rate regimes, the optimized bandwidth allocation, $\varkappa=\varkappa^{\mathrm{opt}}$, requires a number of UAVs that is always less than or equal to that of the equal bandwidth allocation baseline, $\varkappa=1$. This behavior follows directly from the lexicographic Pareto design depicted in Fig.~\ref{sys3}, where maximizing HAPS--RIS user coverage is assigned the highest priority, while UAV reduction is pursued whenever feasible. In the high-rate regime, although the UAV demand increases for all schemes due to stricter rate requirements, the optimized allocation consistently maintains a UAV count that does not exceed that of the equal-allocation baseline, demonstrating a more efficient balance between HAPS--RIS coverage and UAV utilization. In the extreme case $\varkappa=\varkappa_1$, almost the entire bandwidth is allocated to the HAPS--RIS, effectively eliminating the need for the UAV tier from a resource perspective and resulting in zero required UAVs. Finally, as observed in Fig.~\ref{percentageUSerBaselines} and Fig.~\ref{UAVNumberBaselines}, the optimized allocation achieves a better balance between HAPS--RIS user coverage and UAV utilization than the equal bandwidth allocation in both rate regimes. In the proposed framework, the original three-objective formulation in \textbf{OP~1} is relaxed to \textbf{OP~4}, which optimizes only two objectives, while the total average UAV path loss is handled indirectly by minimizing a tractable upper bound derived through the proposed \emph{Theorem 1}. This upper bound remains well controlled across all schemes. In the worst case, the total average UAV path loss is bounded by approximately $108$~dB.

\vspace{-1mm}
\section{Conclusion}
This work establishes HAPS-RIS as a powerful system-level enabler for future UAV-assisted non-terrestrial networks by moving beyond isolated or hierarchical designs toward a fully integrated and jointly optimized architecture. By designing a novel unified multi-objective framework, we reveal how HAPS coverage, UAV scarcity, and UAV path loss are fundamentally intertwined and must be addressed jointly rather than in isolation. A key achievement of this paper is the novel transformation of an intractable large-scale optimization into a scalable and implementable solution. UAV deployment is governed by a k-means clustering equivalence, RIS phase shifts admit a simple closed-form design, and high-dimensional associations collapse into intuitive geometric and bandwidth-level control parameters. The proposed dynamic Pareto optimization method exposes the intrinsic priority structure of integrated aerial networks, showing that fully exploiting HAPS-RIS resources before relying on UAV resources is essential for practical deployment. Extensive simulations uncover clear regime-dependent insights. HAPS-RIS alone is sufficient for low-rate services, while UAV assistance becomes unavoidable as rate demands increase. Moreover, all conventional architectures, i.e., UAV-only, HAPS-RIS-only, and equal bandwidth partitioning, emerge naturally as special cases of the proposed framework through a single bandwidth-portioning factor. The proposed solution consistently outperforms these baselines. Most importantly, the results quantify a tangible trade-off between RIS scale and number of UAVs, enabling designers to replace costly UAV deployments with larger passive surfaces when appropriate, or vice versa. Together, these findings provide not only a unified theoretical framework but also concrete, rate-aware design guidelines for scalable and implementable 6G non-terrestrial networks.


\vspace{-1mm}
\bibliographystyle{IEEEtran}
\bibliography{ref}

\end{document}